\documentclass[journal]{IEEEtran}

\usepackage{amssymb}
\usepackage{latexsym}
\usepackage[numbers]{natbib}
\usepackage{adjustbox}

\usepackage{url}
\usepackage{xcolor}
\usepackage{color,soul}
\usepackage{caption,subcaption,booktabs}

\usepackage{amsmath}
\usepackage{mathtools}
\usepackage{enumitem}

\usepackage{dblfloatfix}
\usepackage{tikz}       % <---
\usepackage{nicematrix} % <---
\usepackage{framed,multirow}

\usepackage{lipsum}
\usepackage[switch]{lineno}
\usepackage{tcolorbox}
\usepackage[colorlinks=true,linkcolor=red]{hyperref}

\usepackage[labelsep=period,labelfont=bf]{caption}
\usepackage[nameinlink]{cleveref}  % must be after hyperef
\crefname{figure}{Fig.}{\textbf{Figure.}}
\crefname{equation}{Eq.}{\textbf{Eq.}}
\crefname{table}{Table}{\textbf{Table.}}
\crefname{section}{Section}{\textbf{Section}}

\definecolor{newcolor}{rgb}{.8,.349,.1}

\hypersetup{
	colorlinks,
	linkcolor=[rgb]{1.0, 0.0, 0.0},
	citecolor=[rgb]{1.0, 0.2, 0.2},
	urlcolor =[rgb]{0.0, 0.0, 0.8}
}

\usepackage[ruled,vlined]{algorithm2e}
\definecolor{commentcolor}{RGB}{110,154,155}   % define comment color
\newcommand{\PyComment}[1]{\ttfamily\textcolor{commentcolor}{\# #1}}  % add a "#" before the input text "#1"
\newcommand{\PyCode}[1]{\ttfamily\textcolor{black}{#1}} % \ttfamily is the code font
\begin{document}
% % make the title area
% \maketitle

\title{MoMA: Momentum Contrastive Learning with Multi-head Attention-based Knowledge Distillation for Histopathology Image Analysis}

\author{~Trinh~Vuong and 
~Jin~Tae~Kwak$^{*}$ \\
\{trinhvg, jkwak\}@korea.ac.kr \\
$*$ Corresponding author \\

% <-this % stops a space
\thanks{Trinh Vuong and Jin Tae Kwak are from School of Electrical Engineering, Korea University, Seoul, Korea}

}

% make the title area
\maketitle

\begin{abstract}
%% Text of abstract
There is no doubt that advanced artificial intelligence models and high quality data are the keys to success in developing computational pathology tools. Although the overall volume of pathology data keeps increasing, a lack of quality data is a common issue when it comes to a specific task due to several reasons including privacy and ethical issues with patient data. In this work, we propose to exploit knowledge distillation, i.e., utilize the existing model to learn a new, target model, to overcome such issues in computational pathology. Specifically, we employ a student-teacher framework to learn a target model from a pre-trained, teacher model without direct access to source data and distill relevant knowledge via momentum contrastive learning with multi-head attention mechanism, which provides consistent and context-aware feature representations.
This enables the target model to assimilate informative representations of the teacher model while seamlessly adapting to the unique nuances of the target data. 
The proposed method is rigorously evaluated across different scenarios where the teacher model was trained on the same, relevant, and irrelevant classification tasks with the target model. Experimental results demonstrate the accuracy and robustness of our approach in transferring knowledge to different domains and tasks, outperforming other related methods. Moreover, the results provide a guideline on the learning strategy for different types of tasks and scenarios in computational pathology. Code is available at: \url{https://github.com/trinhvg/MoMA}.
    
\end{abstract}

\begin{IEEEkeywords}
Knowledge distillation, momentum contrast, multi-head self-attention, computational pathology
\end{IEEEkeywords}

\IEEEpeerreviewmaketitle

%% main text
\section{Introduction} \label{section:intro}

    Computational pathology is an emerging discipline that has recently shown great promise to increase the accuracy and robustness of conventional pathology, leading to improved quality of patient care, treatment, and management \citep{cui2021artificial}. Due to the developments of advanced artificial intelligence (AI) and machine learning (ML) techniques and the availability of high-quality and -resolution datasets, computational pathology approaches have been successfully applied to various aspects of the routine workflow in the conventional pathology from nuclei detection \citep{graham2019hover}, tissue classification \citep{marini2021semi}, and disease stratification \citep{chunduru2022prognostic} to survival analysis \citep{huang2021integration, li2023self}. However, recent studies have pointed out that the issues with the generalizability of computational pathology tools still remain unsolved yet \citep{stacke2020measuring, aubreville2023mitosis}.
     
    %These are, by and large, relying on the recent developments of artificial intelligence (AI) and machine learning (ML) techniques and the availability of high-quality and -resolution datasets. 

    In order to build accurate and reliable computational pathology tools, not only an advanced AI model but also a large amount of quality data is needed. In computational pathology, both the learning capability of the recent AI and ML techniques and the amount of the available pathology datasets keep increasing. However, the quantity of publicly available datasets are far fewer than those in other disciplines such as natural language processing (NLP) \citep{ghorbani2022scaling} and computer vision \citep{zhai2022scaling, dehghani2023scaling}. It is partially due to the nature of pathology datasets, which include multi-gigapixel whole slide images (WSIs) and thus it is hard to share them across the world transparently, but also due to the privacy and ethical issues with patient data. 
    Moreover, the diversity of pathology datasets is limited. For example, Kather19 \citep{kather2019predicting} contains 100,000 image patches of 9 different colorectal tissue types. These 100,000 image patches were initially generated from only 86 images. GLySAC \citep{doan2022sonnet} includes 30,875 nuclei of 3 different cell types from 59 gastric image patches. These were prepared from 8 WSIs that were digitized using a single digital slide scanner. The lack of diversity among scanners in the dataset also hampers the generalization of AI models in computational pathology \citep{stacke2020measuring}.
    In fact, there have been some efforts to provide a large amount of diverse pathology datasets. For instance, PANDA \citep{bulten2022artificial} is a dataset for prostate cancer Gleason grading, which includes 12,625 WSIs from 6 institutes using 3 different digital slide scanners. However, when it comes to a specific computational pathology task, it is still challenging to obtain or have access to a sufficient number of diverse datasets. The collection of such datasets is time- and labor-intensive by any means. Therefore, there is an unmet need for computational pathology in developing task-specific models and tools.

    Transfer learning is one of the most widely used learning methods to overcome the shortage of datasets by re-using or transferring knowledge gained from one problem/task to other problems/tasks. Though it has been successful and widely adopted in pathology image analysis and other disciplines, most of the previous works used the pre-trained weights from natural images such as ImageNet or JFT \citep{hosseinzadeh2021systematic, dosovitskiy2021an}. A recent study demonstrated that the off-the-shelf features learned from these natural images are useful for computational pathology tasks, but the amount of transferable knowledge, i.e., the effectiveness of transfer learning is heavily dependent on the complexity/type of pathology images likely due to differences in image contents and statistics \citep{li2020much}. 
    As the number of publicly available pathology image datasets increases, the pre-trained weights from such pathology image datasets may be used for transfer learning; however, it is unclear whether the quantity is large enough or the variety is diverse enough. It is generally accepted that there are large intra- and inter-variations among pathology images. Hence, the effectiveness of the transfer learning may still vary depending on the characteristics of the datasets. 
    Moreover, the knowledge distillation (KD), proposed by \citep{hinton2015distilling}, is another approach that can overcome the deficit of proper datasets. It not only utilizes the existing model as the pre-trained weight (similar to transfer learning) but also forces the target (student) model to directly learn from the existing (teacher) model, i.e., the student model seeks to mimic the output of the teacher model throughout the training process. Variants of KD have been successfully applied to various tasks, such as model compression \citep{Tian2020Contrastive}, cross-modal knowledge transfer \citep{yuan2022x, ahmed2022cross, zhao2020knowledge}, or ensemble distillation \citep{du2020agree, lin2020ensemble, allen-zhu2023towards}. 
    However, it has not been fully explored for transferring knowledge between models, in particular for pathology image analysis. 

%    Transfer learning has proven to be very useful in many applications. However, it has been also known to be susceptible to adversarial attacks that are transferable from the source model, and thus no prior knowledge of the dataset and annotations being used are required to disrupt the target model [Ahmed], potentially leading to serious privacy and healthcare issues in medicine.  
    
    Herein, we sought to address the question of how to overcome the challenge of limited data and annotations in the field of computational pathology, with the ultimate goal of building computational pathology tools that are applicable to unseen data in an accurate and robust manner. 
    To achieve this goal, we propose an efficient and effective learning framework that can exploit the existing models, built based upon quality source datasets, and learn a target model on a relatively small dataset. The proposed method, so-called \textbf{Mo}mentum contrastive learning with \textbf{M}ulti-head \textbf{A}ttention-based knowledge distillation (\textbf{MoMA}), follows the framework of KD for transferring relevant knowledge from the existing models and adopts momentum contrastive learning and attention mechanism for obtaining consistent, reliable, and context-aware feature representations.
    We evaluate MoMA on multi-tissue pathology datasets under different settings to mimic real-world scenarios in the research and development of computational pathology tools. Compared to other methodologies, MoMA demonstrates superior capabilities in learning a target model for a specific task. Moreover, the experimental results provide a guideline to better transfer knowledge from pre-trained models to student models trained on a limited target dataset. 

    Our main contributions are summarized as follows:
    \begin{itemize}
    \item We develop an efficient and effective learning framework, so-called MoMA, that can exploit the existing models, train on quality datasets, and facilitate building an accurate and robust computational pathology tool on a limited dataset.
    \item We propose to utilize attention-based momentum contrastive learning for KD to transfer knowledge from the existing models to a target model in a consistent and reliable fashion.
    \item We evaluate MoMA on multi-tissue pathology datasets and outperform other related works in learning a target model for a specific task.
    \item We investigate and analyze MoMA and other related works under various settings and provide a guideline on the development of computational pathology tools when limited datasets are available.
    \end{itemize}

\section{Related work}
\subsection{Tissue phenotyping in computational pathology}
    Machine learning has demonstrated its capability to analyze pathology images in various tasks. One of its major applications is tissue phenotyping. In the conventional computational pathology, hand-crafted features, such as color histograms \citep{gorelick2013prostate},  gray level co-occurrence matrix (GLCM) \citep{doyle2012cascaded}, local binary pattern \citep{kather2016multi}, and Gabor filters \citep{doyle2012cascaded, sarkar2017sdl}, have been used to extract and represent useful patterns in pathology images. These hand-crafted features, in combination with machine learning methods such as random forest \citep{paul2015regenerative} and support vector machine \citep{kahya2017classification, nguyen2007multi}, were used to classify types of tissues or grades of cancers.
    With the recent advance in graphics processing unit (GPU) parallel computing, there has been a growing number of deep learning-based methods that achieve competitive results in tissue phenotyping of pathology images. For example, a deep convolutional neural network (CNN) was built and used to detect pathology images with prostate cancer \citep{kwak2011multimodal} and mitosis cells in the breast tissues \citep{li2018deepmitosis}. It was also adopted to detect cancer sub-types; for instance, \citep{le2022prediction} employed a CNN model to identify Epstein-Barr Virus (EBV) positivity in gastric cancers, which is a sign of a better prognosis. To further improve the performance of the deep learning models, various approaches have been proposed, such as multi-task learning \citep{graham2023one}, multi-scale learning \citep{vuong2021multi}, semi-supervised learning \citep{wu2022cross}, or ensemble based models \citep{shi2020graph}.  
    Though these works showed promising results, most of them still suffer from the limited quality of the training datasets, such as a lack of diversity or class imbalance in the datasets \citep{fuchs2011computational}. Transfer learning, which sought to leverage the pre-trained models or weights on other datasets or domains, such as the ImageNet dataset as a starting point, is a simple yet efficient method to alleviate such problems in computational pathology \citep{shinde2021deep, morid2021scoping}.
    
    Although transfer learning with fine-tuning has shown to be effective in many computational pathology applications, this approach does not fully utilize the pre-trained models or weights.
    
. 
    
\subsection{Knowledge distillation}
\label{relatedwork:kd}
    \textit{Knowledge distillation (KD):} KD in deep learning was pioneered by \citep{hinton2015distilling} that transfers knowledge from a powerful source (or teacher) model with large numbers of parameters to another less-parameterized target (or student) model by minimizing the KL divergence between the two models. Transfer learning uses the pre-trained weights from the teacher model as a starting point only. Meanwhile, KD tries to utilize the teacher model throughout the entire training procedure. 
    
    \textit{Feature-Map/Embedding distillation:} Inspired by the vanilla KD \citep{hinton2015distilling}, many variants of KD methods have been proposed, in particular utilizing intermediate feature maps or embeddings.

    For instance, FitNet  \citep{DBLP:journals/corr/RomeroBKCGB14} used \textit{hint} regressions to guide the feature activation of the student model. Attention mechanisms were applied to the intermediate feature maps to improve regression transfer \citep{komodakis2017paying} and to alleviate semantic mismatch across intermediate layers in SemCKD \citep{wang2022semckd}.
    Neuron selectivity transfer (NFT) \citep{huang2017like} proposed to align the distribution of neuron selectivity patterns between student and teacher models. Probabilistic knowledge transfer (PKT) \citep{passalis2020probabilistic} transformed the representations of the student and teacher models into probability distributions and subsequently matched them.

    Moreover, some others sought to transfer knowledge among multiple samples. For example, correlation congruence for KD (CCKD) \citep{peng2019correlation} utilized the correlation among multiple samples for improved knowledge distillation. Contrastive loss, exploiting positive and negative pairs, was also employed for KD \citep{Tian2020Contrastive, xu2020knowledge}.

    Such models have been mainly utilized for model compression \citep{hinton2015distilling}, cross-modal transfer \citep{thoker2019cross}, or ensemble distillation \citep{malinin2019ensemble, lin2020ensemble}. For instance, in DeiT \citep{touvron2021training}, an attention distillation was adopted to distill knowledge from ConvNets teacher model \citep{radosavovic2020designing} and to train vision transformer (ViT) \citep{dosovitskiy2020image} on a small dataset; in \citep{noothout2022knowledge}, a KD method was used to segment chest computed tomography and brain and cardiac magnetic resonance imaging (MRI). In \citep{ahmed2022cross}, a cross-modal KD method was proposed for the knowledge transfer from RGB to depth modality. In \citep{zhao2020knowledge}, knowledge was distilled from a rendered hand pose dataset to a stereo hand pose dataset. These works demonstrate that KD is not only applicable to model compression where the teacher and student models are trained on the same dataset but it could also be used to aid the student model in learning and conducting a relevant task. 
    
    %KD in CPath
    Furthermore, several KD methods have been proposed for computational pathology. In \citep{javed2023knowledge}, a multi-layer feature KD was proposed for breast, colon, and gastrointestinal cancer classification using pathology images. A semi-supervised student-teacher chain \citep{shaw2020teacher, marini2021semi} was proposed to make use of a large unlabeled dataset and to conduct pathology image classification. \citep{hassan2022knowledge} developed another KD method for instance-segmentation in prostate cancer grading. In \citep{dipalma2021resolution}, KD was applied for distilling knowledge across image resolutions where the knowledge from a teacher model, trained on high-resolution images, was distilled to a student model, operating at low-resolution images, to classify celiac disease and lung adenocarcinoma in pathology images. 

    KD has been adopted for different tasks, settings, and problems. In this work, we exploit KD to transfer knowledge between teacher and student models, of which each is built and trained for the same, similar, and different classification tasks. To the best of our knowledge, this is the first attempt to investigate the effectiveness of the KD framework on such stratification of classification tasks in pathology image analysis.

\subsection{Self-supervised momentum contrastive learning}
    % Contrastive learning
    To overcome the lack of (labeled) quality datasets and to improve the model efficiency, several learning approaches have been proposed and explored in the AI community. 
    Self-supervised learning emerges as an approach to learning the feature representation of an input (i.e., a pathology image in this study) in the absence of class labels for the target task. It has been successfully adopted to learn the feature representation in both NLP and computer vision tasks. Utilizing pretext tasks such as rotation \citep{gidaris2018unsupervised}, colorization \citep{goyal2019scaling}, or jigsaw solving \citep{noroozi2016unsupervised} is a popular self-supervised learning approach for computer vision tasks. 
    
    Contrastive learning is another self-supervised learning paradigm that exploits similar and/or dissimilar (contrasting) samples to enhance the representation power of a model, in general, as a pre-training mechanism. 
    Intuitively, in contrastive learning, the model learns to recognize the difference in the feature representations among different images. 
    There are three main variations in contrastive learning, namely end-to-end SimCLR \citep{chen2020simple}, contrastive learning with a memory bank \citep{wu2018unsupervised}, and momentum contrast MoCo \citep{he2020momentum}. End-to-end SimCLR \citep{chen2020simple} has been the most natural setting where the positive and negative representations are from the same batch and updated the model end-to-end by back-propagation, which requires a large batch size. 
    The optimization with a large batch size is challenging \citep{chen2022why}; it is even harder for pathology image analysis since the deep learning models perform better with a large input image, providing more contextual information in high resolution. 
    Constrative learning with a memory bank \citep{wu2018unsupervised} was proposed to store the representations of the entire training samples. For each batch, the additional negative representations were randomly sampled from the memory bank without backpropagation. This approach permits the support of large negative samples without requiring a large volume of GPU memory. 

    In CRD \citep{Tian2020Contrastive}, the memory bank contrastive learning was incorporated into the KD framework to conduct the contrastive representation distillation. However, the feature representation of each sample in the memory bank was updated when it was last seen, i.e., the feature representations were obtained by the encoders at various steps throughout the training procedure, potentially resulting in a high degree of inconsistency. MoCo \citep{he2020momentum} offers smoothly evolving encoders over time; as a result, the negative samples in the memory bank become more consistent.

    Due to its advanced ability to learn the feature representation of an input image without the burden of labeling, both self-supervised learning and contrastive learning have been applied to computational pathology in various tasks \citep{ciga2022self}. ImPash \citep{vuong2023impash} adopted contrastive learning to obtain an encoder with improved consistency and invariance of feature representations, leading to robust colon tissue classification. In \citep{li2022lesion, yang2022cs}, contrastive learning was employed to analyze WSIs without the need for pixel-wise annotations. In \citep{li2022lesion}, an advanced scheme to update the memory bank was proposed to store the features from different types/classes of WSIs, using the WSI-level labels as the pseudo labels for the patches.
    
    In our KD context, teacher and student models are trained on different datasets/tasks/domains. Employing a stationary teacher model would only help if the only purpose of the student model is to exactly mimic the teacher model. Inspired by MoCo, we let the teacher model slowly evolve along with the student model on the target dataset. To further improve MoCo, we propose to incorporate the attention mechanism into MoCo so as to pay more attention to important positive and negative samples. 
    
\subsection{Attention}
    Recently, the attention mechanism, which sought to mimic the cognitive attention of human beings, has appeared as the centerpiece of deep learning models both in computer vision and NLP. Attention in deep learning is, in general, utilized to focus on some parts of images, regions, or sequences that are most relevant to the downstream tasks. 
    There are various kinds of attention mechanisms in deep learning, such as spatial attention \citep{jaderberg2015spatial}, channel attention in squeeze-and-excitation (SE) \citep{hu2018squeeze}, and positional-wise attention \citep{huang2019ccnet}. In \citep{bilal2023aggregation}, a comprehensive review of the existing literature on various types of aggregation methods using the attention module in histopathology image analysis is presented. In \citep{sharma2021cluster}, a weighted-average aggregation technique was employed to aggregate patch-level representations, generating WSI-level representations for breast metastasis and celiac gastric cancer detection. Furthermore, \citep{hashimoto2020multi} introduced a novel approach for blood cancer sub-type classification through domain adversarial attention multi-instance learning, utilizing the attention mechanism proposed in \citep{ilse2018attention}.
    
    Attention has been effectively combined with KD, such as in AT \citep{komodakis2017paying}. This approach used an activation-based spatial attention map, which is generated by averaging the values across the channel dimensions between the teacher and student models. A similar attention approach was proposed for computational pathology in \citep{javed2023knowledge}. Such studies used attention to effectively transfer knowledge between the student and teacher models through multiple intermediate feature maps.
    
    Self-attention, first introduced by \citep{vaswani2017attention}, enables the estimation of the relevance of a particular image, region, or sequence to others within a given context. Self-attention is one of the main components in the recent transformer-based models \citep{khan2022transformers}. 

    Self-attention is not only utilized as a component within the transformer model but it is also employed as an attention module itself. An insightful study in \citep{saldanha2023self} demonstrated that the integration of self-supervised feature extraction into attention-based multiple-instance learning leads to robust generalizability in predicting pan-cancer mutations. In the context of cancer cell detection, \citep{sugimoto2022multi} introduced a modified self-attention mechanism based on concatenation. Another interesting variation of self-attention is co-attention \citep{chen2021multimodal}, which was applied to a multimodal transformer for survival prediction using WSIs and genome data. 
    In our work, we adopt the self-attention module into the momentum contrastive learning framework to learn and utilize the correlation/relevance among the positive pairs and negative pairs.	
	%%%%%%%%%%%%%%%%%%%%%%%%%%%%%%%%%%%%%%%%%%%%%%%%%%%%%%%%%%%%%%%%%%%%%%%%%%%%%%%%%%%%%%%%%%%%
	
\section{Methods} \label{section:methods}

    \begin{figure*}[t!]
        \centering
        \includegraphics[width=\textwidth]{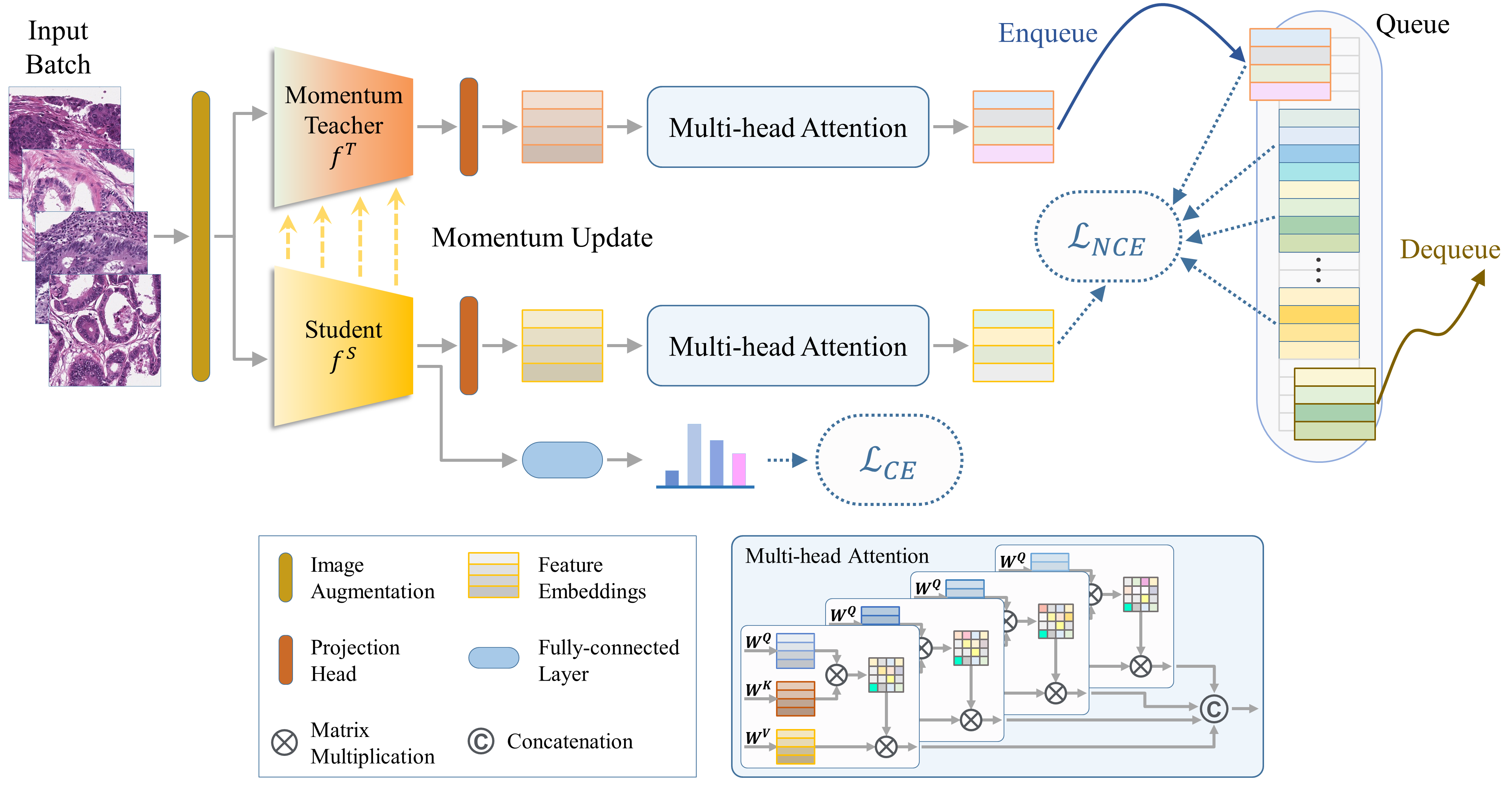}
        \caption{Overview of the MoMA: Attention-Augmented Momentum Contrast Knowledge Distillation framework. A batch of input images is encoded by the student encoder ($f^S$), and the momentum teacher ($f^T$), and each feature representation is re-weighted with regard to other images in the batch as the context. A classifier is added on top of the student encoder. The student model is jointly optimized by contrastive loss and cross-entropy loss.  }
        \label{fig:overview}
    \end{figure*}

\subsection{Problem formulation} \label{section:kd}

The overview of the proposed MoMA is shown in Fig. \ref{fig:overview} and Alg. \ref{algo:algo} in the Appendix A.
Let $D^{SC} = \{(\textbf{x}_i, \textbf{y}_i)\}^{N_{SC}}_{i=1}$ be a source/teacher dataset and $D^{TG} = \{(\textbf{x}_i, (\textbf{y}_i)\}^{N_{TG}}_{i=1}$ be a target/student dataset where  $\textbf{x}_i$ and  $\textbf{y}_i$ represent the $i^{th}$ pathology image and its ground truth label, respectively, and $N_{SC}$ and $N_{TG}$ represent the number of source and target samples ($N_{SC} \gg N_{TG}$), respectively. 
The source/teacher dataset refers to the dataset that is utilized to train a teacher model and the target/student dataset denotes the dataset that is employed to learn a target/student model.
Let $\mathcal{F}^T$ be a teacher model and $\mathcal{F}^S$ be a student model. $\mathcal{F}^T$ consists of a teacher encoder $f^T$ and a teacher classifier $g^T$. $\mathcal{F}^S$ includes a student encoder $f^S$ and a student classifier $g^S$. 
In addition to $\mathcal{F}^T$ and $\mathcal{F}^S$, MoMA includes a teacher projection head ($p^T$), a teacher attention head ($h^T$), a student projection head ($p^S$), and a student attention head ($h^S$). Given an input image $\textbf{x}_i$, $f^T$ and $f^S$ extracts initial feature representations, each of which is subsequently processed by a series of a projection head and an attention head, i.e., $p^T$ followed by $h^T$ or $p^S$ followed by $h^S$, to improve its representation power. $g^T$ and $g^S$ receive the initial feature representations and conduct image classification. $g^T$ is only utilized during the training of $\mathcal{F}^T$.
    
    Due to the restrictions on sharing medical data, we assume a scenario where $\mathcal{F}^T$ has been already trained on $D^{SC}$, the pre-trained weights of $\mathcal{F}^T$ are available, but the direct access to $D^{SC}$ is limited.
    Provided with the pre-trained $f^T$, the objective of MoMA is to learn $\mathcal{F}^S$ on $D^{TG}$ in an accurate and robust manner.
    For optimization, MoMA exploits two learning paradigms: 1) KD framework and 2) momentum contrastive learning. 
    Combining the two learning methodologies, MoMA permits a robust and dynamic transfer of knowledge from $f^T$, which was pre-trained on a high-quality dataset, i.e., $D^{SC}$, to a target $f^S$, which is trained on a limited dataset, i.e., $D^{TG}$.

\subsection{Network architecture} \label{section:arc}
    We construct $f^S$ and $f^T$ using the identical architecture of CNN,  i.e., EfficientNet-b0. 
    Both $p^S$ and $p^T$ are composed of multilayer perceptron (MLP) layers that are composed of a sequence of a fully-connected layer (FC), a ReLU layer, and an FC layer; the resultant output of each projector is a 512-dimensional vector.
    $h^S$ and $h^T$ represent the teacher and student multi-head self-attention layers (MSA) that are described in detail in \ref{section:MSA}. 
    Classifiers ($g^T$ and $g^S$) simply contain a single FC layer.
    
    During training, only $\{f^S, p^S, h^S, h^T\}$ are learned via gradient backpropagation, while we adopt the momentum update policy to update $\{f^{mT}, p^{mT} \}$ using $\{f^{S}, p^{S}\}$ where $f^{mT}$ and $p^{mT}$ are the momentum teacher encoder and projection head, which are described in section \ref{section:moco}.
    
    During inference, we only keep the student encoder $f^S$ and the classifier $g^S$ and discard the momentum teacher encoder $f^{mT}$, the student projection head $p^S$, the teacher projection head $p^{mT}$, the teacher classifier $g^T$, and the multi-head attention layers $\{h^S, h^T\}$. This results in the inference model that is identical to EfficientNet-b0.
   
\subsection{Momentum contrastive learning with multi-head attention}  \label{section:moco_teacher}
\subsubsection{Momentum contrastive learning} \label{section:moco}

In conventional self-supervised contrastive learning, such as in MoCo \citep{chen2020improved}, negative representations are obtained from distinct image pairs, while positive representations are generated from two different views of the same input image. The variations in the positive representations arise from manually curated data augmentation techniques. 
In MoMA, however, the positive representations for an input image are acquired by separately employing the student and teacher models. The variations in the positive representations are ascribable to the difference in the two models, which were trained on separate datasets.

To further advance the feature representation, we adopt the momentum update strategy from MoCo to maintain a consistent dictionary for contrastive learning and introduce a self-attention mechanism (Section \ref{section:MSA}) to focus on important samples within a batch of positive samples and to assign appropriate weights to negative samples before they are added to the memory bank. This strategy enhances the effectiveness of contrastive learning by ensuring that the network prioritizes relevant information and mitigates the impact of dataset differences between the student and teacher networks.

Inspired by MoCo, our MoMA registers a queue of negative representations  $\textbf{Z}^{queue}$ to increase the number of negative samples without high GPU memory demand. In every training iteration, we update  $\textbf{Z}^{queue}$ by enqueuing a new batch of $N_B$ feature representations obtained from the teacher model and dequeuing the oldest $N_B$ feature representations. 
To guarantee consistency among the negative samples in  $\textbf{Z}^{queue}$, we introduce the momentum teacher encoder $f^{mT}$, which is updated along with the student encoder $f^S$ via the momentum update rule, following MoCo-v2 \citep{chen2020improved}. Formally, we denote the parameters of $f^{mT}$ and $p^{mT}$ as $\theta_{mT}$ and those of $f^S$ and $p^S$ as $\theta_S$, we update $\theta_{mT}$ by:
\begin{equation}
\theta_{mT} \leftarrow \alpha \theta_{mT} + (1-\alpha)\theta_S.   
\end{equation}
where $\alpha$ is a momentum coefficient to control the contribution of the new weights from the student model. We empirically set $\alpha$ to 0.9999, which is used in MoCo-v2 \citep{chen2020improved}.

Given a batch of input images $\{x_i\}_{i=1}^{N_B}$, a batch of two feature representations  $\textbf{Z}^{mT}=\{\textbf{z}_i^{mT}\}_{i=1}^{N_B}$ and  $\textbf{Z}^S=\{\textbf{z}_i^S\}_{i=1}^{N_B}$ are obtained as follows:
\begin{align}
\textbf{z}_i^{mT} = h^T(p^{mT}(f^{mT}(\textbf{x}_i))), i=1,...,N_B \label{eq:zs} \\   
\textbf{z}_i^{S} = h^S(p^S(f^{S}(\textbf{x}_i))), i=1,...,N_B \label{eq:zt}.
\end{align} 
$\textbf{Z}^{mT}=\{\textbf{z}_i^{mT}\}_{i=1}^{N_B}$ are used to update   $\textbf{Z}^{queue}=\{\textbf{z}^{queue}_i\}_{i=1}^{N_Q}$ where $N_Q$ is the size of   $\textbf{Z}^{queue}$ ($N_Q=16,384$). 
At each iteration,   $\textbf{Z}^{mT}$ is enqueued into   $\textbf{Z}^{queue}$, and the oldest batch of feature representations are dequeued from   $\textbf{Z}^{queue}$, maintaining a number of recent batches of feature representations. For each input image   $\textbf{x}_i$, a positive pair is defined as ($\textbf{z}^S_i$, $\textbf{z}^T_i$) and a number of negative pairs are defined as   $\{(\textbf{z}^S_i, \textbf{z}^{queue}_j)| j=1,...,N_Q\}$. Then, the objective function forces the positive pair to be closer and the negative pairs to be far apart in an SSCL fashion, which is described in section \ref{section:loss}.
  
\subsubsection{Multi-head attention for augmented feature representation}  \label{section:MSA}
    We adopt the self-attention (SA) mechanism, which was first introduced by \citep{vaswani2017attention}, to reweight the feature representation of an input image with respect to the context of other images in the same iteration/batch. Formally, given a batch of $N_B$ input images $\textbf{X} = \{\textbf{x}_i\}_{i=1}^{N_B} \in \mathbf{R}^{N_B\times C\times H \times W}$, we obtain $N_B$ $d$-dimensional feature embeddings $\textbf{E} = \{\textbf{e}_i\}_{i=1}^{N_B} \in \mathbf{R}^{N_B\times d}$. 

    Using $\textbf{E}$, we define a triplet of learnable weight matrices $W^Q \in \mathbf{R}^{d\times d_q}$, $W^K \in \mathbf{R}^{d\times d_k}$, and $W^V \in \mathbf{R}^{d\times d_v}$ that are used to compute queries $\textbf{Q} = \textbf{E} W^Q \in \mathbf{R}^{N_{B}\times d_q}$, keys $\textbf{K} = \textbf{E} W^K \in \mathbf{R}^{N_{B}\times d_k}$, and values $\textbf{V} = \textbf{E} W^V \in \mathbf{R}^{N_{B}\times d_v}$ where $d_q = d_k = d_v$ denote the dimension of queries, keys, and values, respectively. 

Then, re-weighted feature representations $\textbf{Z} \in \mathbf{R}^{N_B\times d_v}$ are given by, 
    
$$ \textbf{Z} = \operatorname{softmax} \left(\frac{QK^T}{\sqrt{d_q}} \right) \textbf{V}. $$
By applying SA $h$ times and concatenating the output of $h$ SA heads, we obtain the multi-head SA (MSA) feature representations. We set the number of SA heads to $h=4$.
MSA is separately applied to the feature representations obtained from the student and teacher models, producing $\textbf{Z}^S$ and $\textbf{Z}^T$, respectively. The feature representations in $\textbf{Z}^{queue}$ were already re-weighted by MSA. Hence, MSA allows for attending to parts of the student positive samples, teacher positive samples, and the enqueued negative samples differently.

Intuitively, the MSA mechanism augments features within the network and enhances its learning capabilities. It enables each token (in our case, each image) to attend to other tokens within the same batch, capturing dependencies and relationships among them. This attention mechanism facilitates the identification of crucial features and patterns that are useful for accurate classification or prediction.
In the context of MoMA, integrating MSA into the training process allows the student model to focus on salient positive samples while appropriately weighting negative samples before enqueuing them into the memory bank. This approach enables the student model to discern relevant information and prioritize essential samples, contributing to improved generalization and discriminative capabilities.

\subsubsection{Objective function}  \label{section:loss}

The objective function for our MoMA framework is given by:
\begin{equation}
% \small
\mathcal{L} = \mathcal{L}_{CE} + \mathcal{L}_{NCE} + \gamma \mathcal{L}_{KL}
\end{equation}
where $\mathcal{L}_{CE}$, $\mathcal{L}_{NCE}$, and $\mathcal{L}_{KL}$ denote cross-entropy loss, InfoNCE loss \citep{oord2018representation}, and Hinton KD loss, respectively, and $\gamma$ is a binary hyper-parameter ($\gamma=1$ or $0$) to determine whether to include $\mathcal{L}_{KD}$ or not depending on the type of distillation tasks, given by: 
\begin{equation}
\gamma = 
    \begin{cases}
    1 & \text{if student and teacher models conduct the same task} \\
    0 & \text{otherwise}
    \end{cases}
    .
\end{equation}

$\mathcal{L}_{CE}$ is given by:
\begin{equation}
% \small
\mathcal{L}_{CE}(Y, O^S) = \sum_{i=1}^{N_B} \sum^c_{j=1} y_i \log \sigma_j (o_i^S) 
\end{equation}
where $\sigma_j$ is the predicted probability for the $j^{th}$ class computed by the softmax function and $o_i^S$ and $y_i$ denote the logit and ground truth of the $i$th image, respectively.
$\mathcal{L}_{KL}$ denotes KL divergence loss to minimize the difference between the predicted probability distributions given by $\mathcal{F}^S$ and $\mathcal{F}^T$ as follows:

\begin{multline}
% \small
\mathcal{L}_{KL}(O^{mT}, O^S) \\ 
=-\mathcal{T}^2 \sum_{i=1}^{N_B} \sum^c_{j=1} \sigma_j \left(\frac{o_i^{mT}}{\mathcal{T}}\right) \left[\log \sigma_j \left(\frac{o_i^S}{\mathcal{T}}\right) -\log \sigma_j \left(\frac{o_i^{mT}}{\mathcal{T}}\right) \right]
\end{multline}

where $\mathcal{T}$ is a softening temperature ($\mathcal{T}$ = 4).
$\mathcal{L}_{NCE}$ is to optimize momentum contrastive learning in a self-supervised manner. 

Using $\textbf{Z}^S$, $\textbf{Z}^{mT}$, and $\textbf{Z}^{queue}$, $\mathcal{L}_{NCE}$ is calculated as follows:

\begin{multline}
% \small
\mathcal{L}_{NCE}(\textbf{Z}^S, \textbf{Z}^{mT}, \textbf{Z}^{queue}) \\ = \sum_{i=1}^{N_B} \log\frac{\exp(\textbf{z}_i^S \cdot \textbf{z}_i^{mT} /\tau)}{ \exp(\textbf{z}_i^S \cdot \textbf{z}_i^{mT} /\tau) + \sum_{j=1}^{N_Q}\exp(\textbf{z}_i^S \cdot \textbf{z}_j^{queue}/\tau)} 
\end{multline}
where $\tau$ is a temperature hyper-parameter ($\tau=0.07$). By minimizing $\mathcal{L}_{NCE}$, we maximize the mutual information between the positive pairs, i.e., $\textbf{Z}^S$ and $\textbf{Z}^{mT}$, and minimize the similarity between $\textbf{Z}^{mT}$ and negative samples from $\textbf{Z}^{queue}$.

\section{Experiments} \label{section:experiment}
   
    \subsection{Datasets} \label{section:datasets}
    \begin{figure*}[t!]
    \centering
    \includegraphics[width=\textwidth]{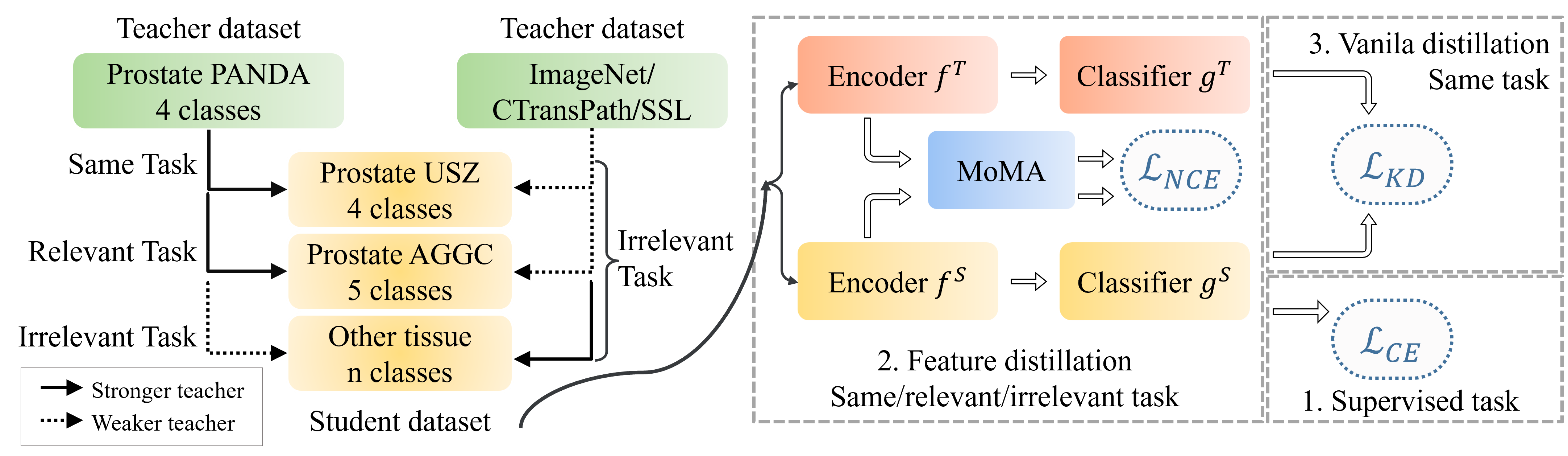}
    \caption{Overview of distillation flow across different tasks and datasets. 1) Supervised task is always conducted, 2) Feature distillation is applied if a well-trained teacher model is available, and 3) Vanilla ${L}_{KD}$ is employed if teacher and student models conduct the same task. SSL stands for self-supervised learning.}
    \label{fig:KD_dataset}
\end{figure*}	
    \subsubsection{Teacher datasets}
    In this study, we employ two (large-scale) teacher datasets, one is a computational pathology dataset, and the other is a natural image dataset. The first one is the Prostate cANcer graDe Assessment (PANDA) dataset \citep{bulten2022artificial}, which includes 5,158 WSIs from Radboud University Medical Center, Netherlands, with pixel-level Gleason grade labels, and 5,456 WSIs from Karolinska Institute, Sweden, with labels such as background, benign tissue, and cancerous tissue. From this, we utilized 5,158 WSIs with Gleason grade annotations, which were digitized them at 20$\times$ magnification using a 3DHistech Pannoramic Flash II 250 scanner ($0.24\mu m \times 0.24\mu m$/pixel) from Radboud University Medical Center. Using the 5,158 WSIs and their pixel-level annotations of benign (BN), grade 3 (G3), grade 4 (G4), and grade 5 (G5), we generated $\sim$100,000 patches of size 512 $\times$ 512 pixels at 20$\times$ magnification. Subsequently, these patches were  randomly divided at the WSI level into a training set with 3,158 WSIs (8,207 BN, 16,015 G3, 24,742 G4, and 4,515 G5), a validation set with 1,000 WSIs (2,602 BN, 4,933 G3, 8,244 G4, and 1,244 G5), and a test set with 1,000 WSIs (2,613 BN, 5,036 G3, 8,809 G4, and 1,239 G5). The second teacher dataset is the well-known ImageNet dataset, which is irrelevant to pathology images and tasks. We note that once the teacher models are trained on each of the teacher datasets, the teacher models are not re-trained on the target dataset; the pre-trained weights from the PyTorch library are adopted for the ImageNet teacher models.

    \subsubsection{Prostate cancer 4-class dataset} 
    \textbf{Prostate USZ} \citep{arvaniti2018automated} was obtained from the Harvard dataverse (https://dataverse.harvard.edu/). It is composed of 886 tissue core images, digitized at 40x magnification, that were scanned by a NanoZoomer-XR Digital slide scanner (Hamamatsu) ($0.23\mu m \times 0.23\mu m$/ pixel) from University Hospital Zurich (USZ). Prostate USZ is extracted at a size of 750 $\times$ 750 pixels. Prostate USZ is used as training (2076 BN, 6303 G3, 4541 G4, and 2383 G5 patches), validation (666 BN, 923 G3, 573 G4, and 320 G5 patches), and test (127 BN, 1602 G3, 2121 G4, and 387 G5 patches) sets for prostate cancer 4-class classification. 
    
    \textbf{Prostate UBC} \citep{nir2018automatic} was acquired from the training set of the Gleason2019 challenge (https://gleason2019.grand-challenge.org/). Prostate Gleason19 is used as an independent test set for prostate cancer 4-class classification. This involves a set of 244 prostate tissue cores that were digitized at 40x magnification  ($0.25\mu m \times 0.25\mu m$/ pixel) using an Aperio digital slide scanner (Leica Biosystems) and annotated by 6 pathologists at the Vancouver Prostate Centre. There are 17,066 image patches (1284 BN, 5852 grade 3, 9682 grade 4, and 248 grade 5), of which each has a size of 690 $\times$ 690 pixels.

    \subsubsection{Prostate cancer 5-class dataset}
    \textbf{
    Prostate AGGC} was obtained from the training set of the Automated Gleason Grading Challenge 2022 (https://aggc22.grand-challenge.org/). The dataset consists of three distinct subsets, all available at 20$\times$ ($0.5 \mu m \times 0.5 \mu m$/pixel). The first subset comprises 105 whole mount images that were scanned using an Akoya Biosciences scanner. From these 105 images, we obtained 133,246 patches including 17,269 stroma, 15,443 BN, 36,627 G3, 57,578 G4, and 6,329 G5 patches of size 512$\times$512 pixels. We utilize these image patches to conduct a 5-fold cross-validation experiment, which is designated as AGGC CV. 
    The second subset consists of 37 biopsy images that were scanned using an Akoya Biosciences scanner. The third subset encompasses 144 whole mount images scanned using multiple scanners from different manufacturers, including Akoya Biosciences (26 images), Olympus (25 images), Zeiss (15 images), Leica (26 images), KFBio (26 images), and Philips (26 images). We reserve the second and third subsets for testing purposes (AGGC test). AGGC test contains 190,451 patches of size 512$\times$512 pixels same size as the  AGGC CV, including  29,225 stroma, 16,272 BN, 53,602 G3, 90,823 G4, and 529 G5 patches. 

    \subsubsection{Colon tissue type classification datasets}
    \textbf{Colon K19} \citep{kather2019predicting} dataset includes 100,000 patches of size 224 $\times$ 244 pixels digitized at 20 $\times$ magnification ($0.5 \mu m \times 0.5 \mu m$/pixel). The patches are categorized into 9 tissue classes: adipose (Ad), background (Bk), debris (De), lymphocytes (Ly), mucus (Mc), smooth muscle (Ms), normal colon mucosa (No), cancer-associated stroma (St), tumor epithelium (Tu). We utilize Colon K19 for the training and validation of colon tissue type classification.

    \textbf{Colon K16} \citep{kather2016multi} contains 5,000 image patches of size 224 $\times$ 244 pixels scanned at 20 $\times$ magnification, ($0.495 \mu m \times 0.495 \mu m$/pixel). There are eight tissue phenotypes, namely tumor epithelium (Tu), simple stroma (St), complex stroma (Complex St), lymphocyte (Ly), debris (De), normal mucosal glands (Mu), adipose (Ad), and background (Bk). The dataset is balanced with 625 patches per class. We use this Colon K16 to test the model that was trained and validated on Colon K19. Since there is no complex stroma in the training set, we exclude this tissue type, resulting in the testing set including 4375 images of 7 tissue classes.

    To resolve the difference in the class label between colon K19 and colon K16, we re-group the 9 classes of colon K19 into 5 classes and 7 classes (excluded complex stroma) into 5 classes, following \citep{abbet2022self}. Specifically, we exclude Complex St and group stroma/muscle and debris/mucus as stroma and debris, respectively. The model is trained on K19 using 9 classes; grouping is only used for inference on K16 purposes. There exist two versions for Colon K19 with and without Macenko stain normalization (SN) while K16 is available without stain normalization; we use Macenko SN to construct the SN version of K16. We use both versions separately for training, validation, and testing purposes.    
    
    \subsubsection{Breast carcinoma sub-type classification dataset} 
    \textbf{BRACS}  \citep{brancati2022bracs}  dataset comprises 547 WSIs scanned using an Aperio AT2 scanner at 40$\times$ magnification ($0.25 \mu m \times 0.25 \mu m$/pixel). The dataset includes 4,539 image patches with varying sizes in a range from 300 $\times$ 300 to 10,000 $\times$ 10,000 pixels. These images are annotated with normal (No), pathological benign (Pb), usual ductal hyperplasia (Ud), flat epithelial atypia (Fe), atypical ductal hyperplasia (Ad), ductal carcinoma in situ (Dc), and invasive carcinoma (Ic). The entire images are divided into a training set (357 No, 714 Pb, 369 Ud, 624 Fe, 387 Ad, 665 Dc, and 521 Ic), a validation set (46 No, 43 Pb, 46 Ud, 49 Fe, 41 Ad, 40 Dc, and 47 Ic), and a test set (81 No, 79 Pb, 82 Ud, 83 Fe, 79 Ad, 85Dc, and 81 Ic).

    \subsubsection{Gastric microsatellite instability classification dataset}
    \textbf{TCGA-STAD} dataset includes 302 WSIs, originally obtained from the TCGA database (https://www.cancer.gov/ccg/research/genome-sequencing/tcga), that are categorized into either microsatellite unstable/highly mutated (MSI) or microsatellite stable (MSS). Following \citep{kather2019deep}, a number of image patches are generated from the 302 WSIs, including a training set of 35 MSI WSIs (50,285 image patches) and 164 MSS WSIs (50,285 image patches) and a test set with 25 MSI WSIs (27,904 image patches) and 78 MSS WSIs (90,104 image patches). All the patches are generated at a size of 224 $\times$ 244 pixels scanned at 20$\times$ magnification ($0.5 \mu m \times 0.5 \mu m$/pixel).

\subsection{Experimental design} \label{section:comparative_experiments}
In order to evaluate the effectiveness of MoMA, we conduct three types of distillation tasks: 1) same task distillation: distillation between prostate cancer classification models, 2) relevant task distillation: distillation from 4-class prostate cancer classification to 5-class prostate cancer classification, and 3) irrelevant task distillation: distillation from prostate cancer classification to colon tissue type classification and to breast carcinoma sub-type classification. 
Fig. \ref{fig:KD_dataset} illustrates the distillation flow and the associated datasets and models.

We also compare MoMA with three different types of competing methods: 
\begin{itemize}
    \item \textbf{Transfer Learning} ($TL$): 1) TC$_{PANDA}$: $f^T$ trained on PANDA without fine-tuning on student datasets, 2) FT$_{None}$: $f^S$ without pre-trained weights, 3) FT$_{ImageNet}$: $f^S$ with pre-trained weights on ImageNet, and 4) FT$_{PANDA}$: $f^S$ with pre-trained weights on PANDA where TC refers to a teacher model and FT denotes fine-tuning on a target (student) dataset.
    \item \textbf{Logits distillation} ($LD$): 1) Vanilla KD \citep{hinton2015distilling}: $f^S$ with vanilla KD method and 2) SimKD \citep{chen2022knowledge}: $f^S$ with re-used $g^T$ but no vanilla KD. Vanilla KD and SimKD are only applied to the same task distillation experiments due to the usage of $g^T$.
    \item \textbf{Feature-Map/Embedding Distillation} ($FD$):  1) FitNet \citep{romero2014fitnets}, 2) AT \citep{komodakis2017paying}, 3) SemCKD \citep{wang2022semckd}, 4) CC \citep{peng2019correlation}, and 5) CRD \citep{Tian2020Contrastive}.  
\end{itemize}

    Moreover, we carry out an ablation study without utilizing MSA (MoMA w/o MSA) to investigate the influence of MSA on strengthening feature representations. 
    We conduct a comparative study to assess the effect of pre-trained weights on the classification performance by employing two sets of pathology-specific pre-trained weights. 
    To further validate the effectiveness of MoMA, we perform a WSI-level classification for MSI.

    \subsection{Implementation Details} \label{section:implementation}

    \subsubsection{Data augmentation:}    
    We employ \textit{RandAugment} \citep{cubuk2020randaugment} for training all the student models. Prostate USZ is resized to $512 \times 512$ pixels during training and testing. Prostate UBC is center cropped from $512 \times 512$ pixels to $448 \times 448$ pixels. Colon K19 is trained and validated at its original size of $224 \times 224$ pixels, and Colon K16 is resized to $224 \times 224$ pixels during inference.
    
    \subsubsection{Training details} 
    
    The teacher model is initialized with the pre-trained weights on ImageNet and then fine-tuned for 100 epochs on PANDA, resulting in TC$_{PANDA}$. Each student model undergoes fine-tuning for 50 epochs on their respective target dataset by using transfer learning or a distillation method with and without pre-trained weights.

    In patch-level classification, except for the prostate AGGC dataset, which uses 5-fold cross-validation with test set results reported, experiments for the prostate USZ, colon classification, and breast classification datasets were conducted five times. The test set results for these datasets are reported as the average across these five independent runs.
    
    For the gastric microsatellite instability WSI-level classification, we randomly partitioned the training set into 5 folds for cross-validation and reported the results on the test set. All models are trained on the image patches using the WSI-level label in a weakly supervised manner. The patch-level classification results are summarized via majority voting to obtain the WSI-level prediction.

    All networks are trained using the Adam optimizer with default parameter values ($\beta_1 =0.9$, $\beta_2 = 0.9999$, $\epsilon= 1.0e^8$), employing a batch size of 64 for prostate datasets and 256 for colon, breast, and gastric datasets. Cross-entropy is utilized as the classifier loss function for all models. All models are implemented using the PyTorch platform and executed on a workstation equipped with two RTX A6000 GPUs.

    \subsection{Quantitative evaluation} \label{section:performance_evaluation}
    We evaluate MoMA and its competing models on the three distillation tasks using 1) Accuracy (ACC), 2) Macro-average F1 (F1), and 3) quadratic weighted kappa ($\kappa_w$). For the relevant distillation task on Prostate AGGC, we use weighted-average F1, F1$_w$ = 0.25 * F1$_{G3}$ + 0.25 * F1$_{G4}$ +0.25 * F1$_{G5}$ +0.125 * F1$_{Normal}$ +0.125 * F1$_{Stroma}$, which is the evaluation metric in the AGGC challenge.

\section{Experimental results} \label{section:expandresults}
\subsection{Same task distillation: prostate cancer classification} \label{section:colon_tma_results}

    \begin{table*}[!t]
    \begin{center}
    \caption{Results of same task distillation. KL denotes the use of KL divergence loss.}
    \label{table:prostate_tma}
    \setlength{\tabcolsep}{2pt} % Default value: 6pt
	\renewcommand{\arraystretch}{1} % Default value: 1
	\begin{adjustbox}{width=\textwidth}
	\begin{tabular}{c|c|ccc|ccc}
% \cline{3-8}
\cline{3-8}
 
     % \toprule
\multicolumn{2}{c}{} & & Prostate USZ (Test I) & &	& Prostate UBC (Test II)	 &	 \\
\toprule
Method &	Pretrained &	ACC($\%$) &	F1 &	$\kappa_w$ &	ACC($\%$) &	F1 &	$\kappa_w$ \\
     \toprule

TC$_{PANDA}$     & 	ImageNet & 	$ 63.4 $ & 	$ 0.526 $ & 	$ 0.531 $ & 	$ 78.2 $ & 	$ 0.580 $ & 	$ 0.680 $ \\ 
FT               & 	None     & 	$ 66.4 \pm 1.6 $ & 	$ 0.566 \pm 0.012 $ & 	$ 0.551 \pm 0.020 $ & 	$ 31.7 \pm 9.6 $ & 	$ 0.239 \pm 0.073 $ & 	$ 0.143 \pm 0.104 $ \\ 
FT & 	ImageNet             & 	$ 67.0 \pm 2.6 $ & 	$ 0.612 \pm 0.027 $ & 	$ 0.604 \pm 0.016 $ & 	$ 71.0 \pm 2.8 $ & 	$ 0.592 \pm 0.026 $ & 	$ 0.619 \pm 0.036 $ \\ 
FT & 	PANDA                & 	$ 72.7 \pm 1.1 $ & 	$ \textbf{0.687} \pm \textbf{0.009} $ & 	$ \textbf{0.671} \pm \textbf{0.005} $ & 	$ 73.1 \pm 1.9 $ & 	$ 0.599 \pm 0.023 $ & 	$ 0.654 \pm 0.031 $ \\ 
\midrule
FitNet  \cite{DBLP:journals/corr/RomeroBKCGB14} & 	PANDA & 	$ 65.7 \pm 3.6 $ & 	$ 0.574 \pm 0.048 $ & 	$ 0.559 \pm 0.056 $ & 	$ 34.5 \pm 19.5 $ & 	$ 0.260 \pm 0.150 $ & 	$ 0.139 \pm 0.131 $ \\ 
AT \citep{komodakis2017paying} & 	PANDA & 	$ 71.2 \pm 1.6 $ & 	$ 0.653 \pm 0.021 $ & 	$ 0.652 \pm 0.023 $ & 	$ 76.0 \pm 3.5 $ & 	$ 0.628 \pm 0.038 $ & 	$ 0.660 \pm 0.053 $ \\ 
CC \citep{peng2019correlation} & 	PANDA & 	$ 69.4 \pm 1.4 $ & 	$ 0.624 \pm 0.016 $ & 	$ 0.608 \pm 0.026 $ & 	$ 51.9 \pm 12.6 $ & 	$ 0.392 \pm 0.100 $ & 	$ 0.268 \pm 0.179 $ \\ 
CRD \citep{Tian2020Contrastive} & 	PANDA & 	$ 70.9 \pm 0.7 $ & 	$ 0.642 \pm 0.012 $ & 	$ 0.639 \pm 0.015 $ & 	$ 70.7 \pm 3.5 $  & 	$ 0.577 \pm 0.032 $ & 	$ 0.610 \pm 0.046 $ \\ 
SemCKD \citep{wang2022semckd}  & 	PANDA & 	$ 69.8 \pm 1.0 $ & 	$ 0.638 \pm 0.013 $ & 	$ 0.635 \pm 0.007 $ & 	$ 75.4 \pm 1.7 $  & 	$ 0.627 \pm 0.009 $ & 	$ 0.685 \pm 0.017 $ \\ 
MoMA (Ours) & 	ImageNet  & 	$ 67.3 \pm 3.4 $ & 	$ 0.617 \pm 0.013 $ & 	$ 0.594 \pm 0.028 $ & 	$ 65.4 \pm 5.8 $ & 	$ 0.534 \pm 0.059 $ & 	$ 0.533 \pm 0.128 $ \\ 
MoMA (Ours) & 	PANDA & 	$ \textbf{73.6} \pm \textbf{1.0} $ & 	$ 0.687 \pm 0.011 $ & 	$ 0.670 \pm 0.010 $ & 	$ 75.5 \pm 1.8 $ & 	$ 0.622 \pm 0.019 $ & 	$ 0.666 \pm 0.015 $ \\ 
\midrule
Vanilla KD \citep{hinton2015distilling}                    & 	PANDA & 	$ 69.2 \pm 0.4 $ & 	$ 0.596 \pm 0.008 $ & 	$ 0.593 \pm 0.005 $ & 	$ 82.9 \pm 0.5 $ & 	$ 0.649 \pm 0.013 $ & 	$ 0.74 \pm 0.013 $ \\ 
SimKD \citep{chen2022knowledge} & 	PANDA & 	$ 65.9 \pm 0.3 $ & 	$ 0.420 \pm 0.001 $ & 	$ 0.413 \pm 0.004 $ & 	$ 79.8 \pm 0.1 $ & 	$ 0.586 \pm 0.001 $ & 	$ 0.673 \pm 0.002 $ \\ 
KL+FitNet \citep{DBLP:journals/corr/RomeroBKCGB14} & 	PANDA & 	$ 68.7 \pm 1.4 $ & 	$ 0.583 \pm 0.030 $ & 	$ 0.585 \pm 0.031 $ & 	$ 61.2 \pm 27.0 $ & $ 0.452 \pm 0.211 $ & 	$ 0.437 \pm 0.279 $ \\ 
KL+AT \citep{komodakis2017paying} & 	PANDA & 	$ 68.7 \pm 1.1 $ & 	$ 0.586 \pm 0.017 $ & 	$ 0.593 \pm 0.014 $ & 	$ 82.0 \pm 1.1 $ & 	$ 0.650 \pm 0.009 $ & 	$ 0.728 \pm 0.018 $ \\ 
KL+CC \citep{peng2019correlation} & 	PANDA & 	$ 68.6 \pm 0.7 $ & 	$ 0.586 \pm 0.016 $ & 	$ 0.589 \pm 0.013 $ & 	$ 81.8 \pm 0.9 $ & 	$ 0.630 \pm 0.015 $ & 	$ 0.725 \pm 0.017 $ \\
KL+CRD \citep{Tian2020Contrastive} & 	PANDA & $ 69.2 \pm 1.0 $ & 	$ 0.595 \pm 0.020 $ & 	$ 0.586 \pm 0.010 $ & 	$ 82.3 \pm 1.7 $ & 	$ 0.640 \pm 0.016 $ & 	$ 0.728 \pm 0.036 $ \\ 
KL+SemCKD \citep{wang2022semckd} & 	PANDA & 	$ 68.9 \pm 0.4 $ & 	$ 0.595 \pm 0.004 $ & 	$ 0.596 \pm 0.002 $ & 	$ 81.9 \pm 0.8 $ & 	$ \textbf{0.652} \pm \textbf{0.014} $ & 	$ 0.734 \pm 0.010 $ \\ 
KL+MoMA (Ours) & 	ImageNet & 	$ 70.3 \pm 0.6 $ & 	$ 0.640 \pm 0.019 $ & 	$ 0.618 \pm 0.011 $ & 	$ 72.9 \pm 3.5 $ & 	$ 0.585 \pm 0.035 $ & 	$ 0.626 \pm 0.027 $ \\
KL+MoMA (Ours) & 	PANDA & 	$ 72.2 \pm 0.6 $ & 	$ 0.638 \pm 0.018 $ & 	$ 0.618 \pm 0.014 $ & 	$ \textbf{83.3} \pm \textbf{0.8} $ & 	$ 0.648 \pm 0.016 $ & 	$ \textbf{0.763} \pm \textbf{0.009} $ \\

    \bottomrule
    \end{tabular}
    \end{adjustbox}
    \end{center}
	\end{table*}

\begin{figure*}[!t]
\centering
\includegraphics[width=\textwidth]{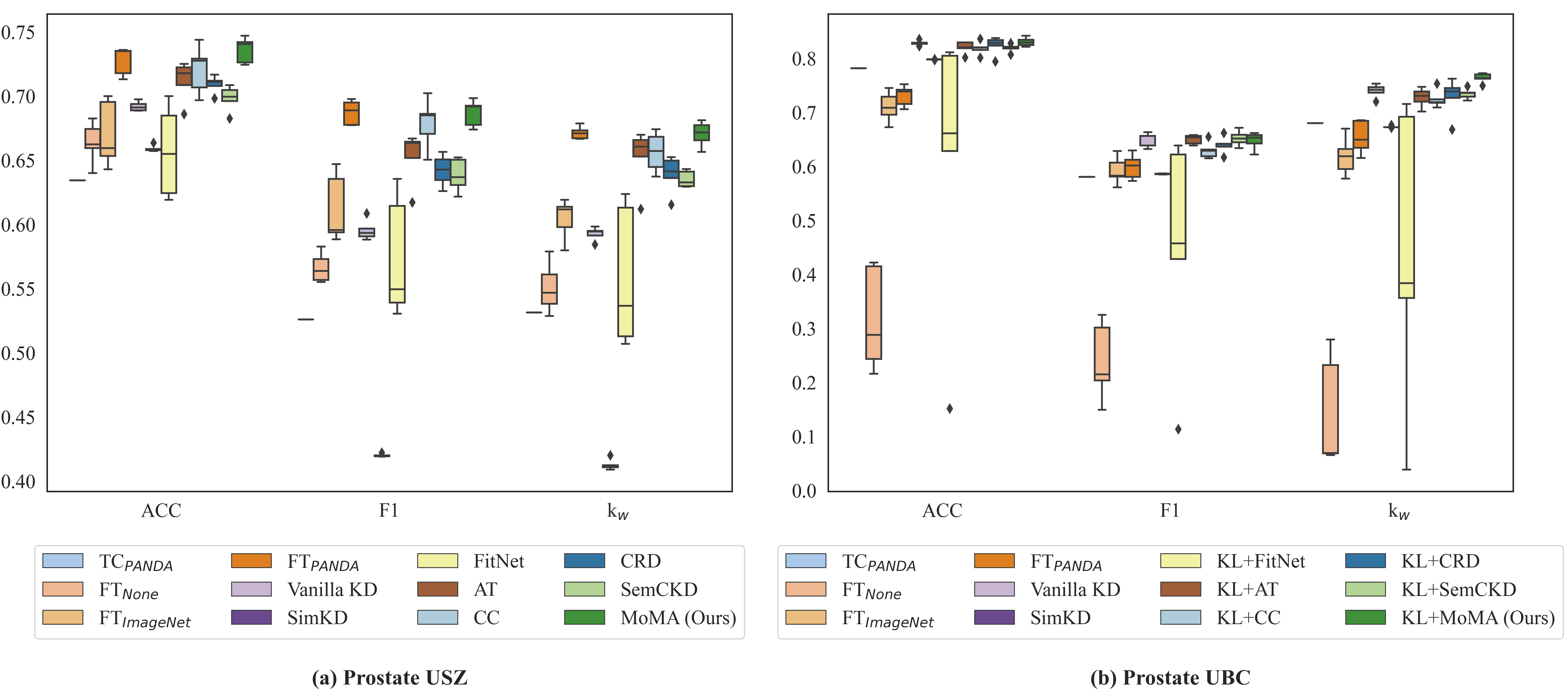}
\caption{Box plots for same task distillation: All the KD models utilize the pre-trained weights from PANDA.}
\label{fig:prostate_tma}
\end{figure*}

Table \ref{table:prostate_tma} and Fig. \ref{fig:prostate_tma}  (and Fig. \ref{fig:confusion_matric_number_prostate_1} and Fig. \ref{fig:confusion_matric_number_prostate_2} in the Appendix C) show the results of MoMA and its competitors on the two TMA prostate datasets (Prostate USZ and Prostate UBC).
On Prostate USZ, the teacher model TC$_{PANDA}$, which was trained on PANDA only, achieved 63.4 $\%$ ACC, 0.526 F1, and 0.531 $\kappa_w$, which is substantially lower to other student models with $TL$, $LD$, and $FD$. 
Among the student models with $TL$, the student model with no pre-trained weights (FT$_{None}$) was inferior to the other two student models; the student model pre-trained on PANDA (FT$_{PANDA}$) outperformed the student model pre-trained on ImageNet (FT$_{ImageNet}$). These indicate the importance of pre-trained weights and fine-tuning on the target dataset, i.e., Prostate USZ.
As for the KD approaches, MoMA$_{PANDA}$, pre-trained on PANDA, outperformed all other KD methods, achieving ACC of 73.6 \%, which is 0.9 \% higher than FT$_{PANDA}$, and F1 of 0.687 and $\kappa_w$ of 0.670, which are comparable to those of FT$_{PANDA}$. 

On the independent test set, Prostate UBC, it is remarkable that TC$_{PANDA}$ achieved 78.2 $\%$ ACC and 0.680 $\kappa_w$, which are superior to those of all the student models with $TL$, likely suggesting that the characteristic of PANDA is more similar to Prostate UBC than Prostate USZ. 
The performance of the student models with $TL$ and $FD$ was similar to each other between Prostate USZ and Prostate UBC; for instance, MoMA$_{PANDA}$ obtained higher ACC but lower F1 and $\kappa_w$ on Prostate UBC than on Prostate USZ.
As MoMA and other student models with $FD$ adopt vanilla KD by setting $\gamma$ to 1 in $\mathcal{L}$, i.e., mimicking the output logits of the teacher model, there was, in general, a substantial increase in the performance on Prostate UBC. MoMA$_{PANDA}$, in particular, achieved the highest ACC of 83.3 \% and $\kappa_w$ of 0.763 overall models under consideration, which are 11.1 \% and 0.145 higher than those on Prostae USZ in ACC and $\kappa_w$, respectively.

By randomly sampling 25\% and 50\% of the training set, we repeated the above experiments using MoMA and other competing models to assess the effect of the size of the training set. The results of the same task distillation using 25\% and 50\% of the training set are available in the Appendix B (Table \ref{table:prostate_tma_50} and \ref{table:prostate_tma_25}). 
The experimental results were more or less the same as those using the entire training set. MoMA$_{PANDA}$ was comparable to FT$_{PANDA}$ on Prostate USZ. KL+MoMA$_{PANDA}$ outperformed the competing models on Prostate UBC. These results validate the effectiveness of MoMA on the extremely small target dataset.

\subsection{Relevant task distillation: prostate cancer classification} \label{section:Relevant_task}
\begin{table*}[!t]
\begin{center}

\caption{Results of relevant task distillation.}
\label{table:prostate_aggc}

\setlength{\tabcolsep}{2pt} % Default value: 6pt
\renewcommand{\arraystretch}{1} % Default value: 1
\begin{adjustbox}{width=\textwidth}
\begin{tabular}{c|c|ccc|ccc}
\cline{3-8}

\multicolumn{2}{c}{} & &  AGGC CV (Test I)	 &  &	& AGGC test (Test II)	 & 	 \\
\toprule
Method &	Pretrained &	ACC($\%$) &	F1$_w$ &	$\kappa_w$ &	ACC($\%$) &	F1$_w$ &	$\kappa_w$ \\
\toprule
FT & 	None  & 	$ 74.0 \pm 3.7 $ & 	$ 0.644 \pm 0.039 $ & 	$ 0.774 \pm 0.043 $ & 	$ 62.8 \pm 1.0 $ & 	$ 0.483 \pm 0.021 $ & 	$ 0.483 \pm 0.089 $ \\ 
FT & 	ImageNet & 	$ 73.8 \pm 4.5 $ & 	$ 0.645 \pm 0.057 $ & 	$ 0.764 \pm 0.043 $ & 	$ 76.0 \pm 2.1 $ & 	$ 0.593 \pm 0.012 $ & 	$ 0.770 \pm 0.042 $ \\ 
FT & 	PANDA & 	$ 76.6 \pm 2.8 $ & 	$ 0.660 \pm 0.025 $ & 	$ 0.792 \pm 0.031 $ & 	$ 76.5 \pm 1.9 $ & 	$ 0.603 \pm 0.013 $ & 	$ 0.771 \pm 0.036 $ \\ 
\midrule
FitNet \citep{DBLP:journals/corr/RomeroBKCGB14} & 	PANDA & 	$ 74.2 \pm 4.5 $ & 	$ 0.638 \pm 0.050 $ & 	$ 0.774 \pm 0.037 $ & 	$ 69.6 \pm 7.6 $ & 	$ 0.554 \pm 0.044 $ & 	$ 0.653 \pm 0.095 $ \\ 
AT \citep{komodakis2017paying} & PANDA &	$ 75.3 \pm 3.0 $ & 	$ 0.649 \pm 0.053 $ & 	$ 0.782 \pm 0.046 $ & 	$ 74.3 \pm 3.4 $ & 	$ 0.583 \pm 0.017 $ & 	$ 0.779 \pm 0.038 $ \\ 
CC \citep{peng2019correlation} & PANDA &	$ \textbf{77.4} \pm \textbf{3.0} $ & 	$ 0.668 \pm 0.352 $ & 	$ 0.796 \pm 0.350 $ & 	$ 77.2 \pm 1.5 $ & 	$ 0.603 \pm 0.016 $ & 	$ 0.760 \pm 0.007 $ \\ 

CRD  \citep{Tian2020Contrastive} &PANDA & 	$ 76.7 \pm 2.7 $ & 	$ 0.663 \pm 0.040 $ & 	$ 0.796 \pm 0.036 $ & 	$ \textbf{77.3} \pm \textbf{4.0} $ & 	$ 0.602 \pm 0.033 $ & 	$ 0.789 \pm 0.029 $ \\ 
SemCKD \citep{wang2022semckd} & 	PANDA & 	$ 74.1 \pm 5.8 $ & 	$ 0.639 \pm 0.043 $ & 	$ 0.762 \pm 0.058 $ & 	$ 65.7 \pm 13.3 $ & 	$ 0.532 \pm 0.087 $ & 	$ 0.642 \pm 0.107 $ \\ 
MoMA (Ours) & 	ImageNet & 	$ 75.7 \pm 4.8 $ & 	$ 0.663 \pm 0.054 $ & 	$ 0.772 \pm 0.035 $ & 	$ 75.5 \pm 3.5 $ & 	$ 0.587 \pm 0.019 $ & 	$ 0.788 \pm 0.021 $ \\ 
MoMA (Ours)& 	PANDA & 	$ 77.1 \pm 2.0 $ & 	$ \textbf{0.670} \pm \textbf{0.042} $ & 	$ \textbf{0.798} \pm \textbf{0.029} $ & 	$ 77.1 \pm 2.3 $ & 	$ \textbf{0.609} \pm \textbf{0.015} $ & 	$ \textbf{0.794} \pm \textbf{0.018} $ \\ 

\bottomrule
\end{tabular}
\end{adjustbox}
\end{center}
\end{table*}

 \begin{figure*}[!t]
    \centering
    \includegraphics[width=\textwidth]{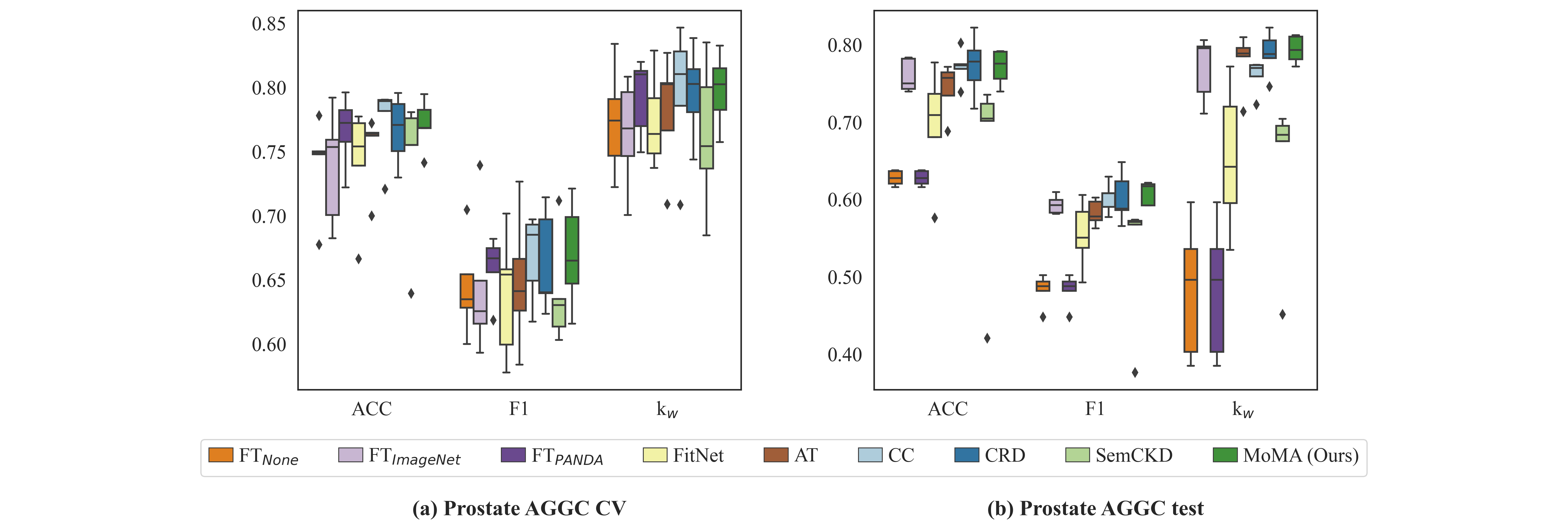}
    \caption{Box plots for relevant task distillation. All the KD models utilize the pre-trained weights from PANDA.}
    \label{fig:prostate_wsi}
\end{figure*}

Table \ref{table:prostate_aggc} and Fig. \ref{fig:prostate_wsi} (and Fig. \ref{fig:confusion_matric_number_aggc_1} and Fig. \ref{fig:confusion_matric_number_aggc_2} in the Appendix C) show the results of MoMA and its competing methods on relevant task distillation, i.e., distillation from 4-class prostate cancer classification to 5-class prostate cancer classification (Prostate AGGC).
The two tasks share 4 classes in common, and thus the direct application of the teacher model and logits distillation is infeasible.
In the cross-validation experiments (AGGC CV), MoMA$_{PANDA}$, on average, achieved the best F1$_w$ of 0.670 and $\kappa_w$ of 0.798 and obtained the second best ACC of 77.1 \%. The performance of FT$_{PANDA}$ was generally comparable to the student models with $FD$. Other student models with $TL$ were, by and large, inferior to the ones with $FD$.
In a head-to-head comparison between the Prostate AGGC CV and Prostate AGGC test, there was, in general, a slight performance drop, likely due to the differences in the type of images and scanners. Though there was a performance drop, similar trends were found across different models between AGGC CV and AGGC test. We also note that MoMA$_{PNADA}$, on average, was superior to all the competitors on two evaluation metrics (F1$_w$ and $\kappa_w$) and attained the third best ACC, which is 0.2 \% lower than CRD.

\subsection{Irrelevant task distillation: colon tissue type classification} \label{section:prostate_tma_results}

    \begin{table*}[!t]
    \begin{center}
    \caption{Results of irrelevant task distillation: colon tissue type classification.}
    \label{table:colon_result}
    \setlength{\tabcolsep}{2pt} % Default value: 6pt
	\renewcommand{\arraystretch}{1} % Default value: 1
	\begin{adjustbox}{width=\textwidth}
	\begin{tabular}{c|c|ccc|ccc}
% \cline{3-8}
     \toprule
\multicolumn{2}{c}{} & & Colon K16 SN (Test I) & &	& Colon K16 (Test II) 	 &	 \\

\midrule
Method &	Pretrained &	ACC ($\%$)  &	F1 &	$\kappa_w$ &	ACC ($\%$)  &	F1 &	$\kappa_w$ \\

\midrule
FT & 	None  & 	$ 76.8 \pm 4.7 $ & 	$ 0.752 \pm 0.057 $ & 	$ 0.871 \pm 0.021 $ & 	$ 81.6 \pm 1.9 $ & 	$ 0.814 \pm 0.020 $ & 	$ 0.858 \pm 0.011 $ \\ 
FT & 	ImageNet & 	$ 83.8 \pm 0.6 $ & 	$ 0.834 \pm 0.007 $ & 	$ 0.879 \pm 0.007 $ & 	$ 86.2 \pm 0.7 $ & 	$ 0.861 \pm 0.008 $ & 	$ 0.898 \pm 0.011 $ \\ 
FT & 	PANDA & 	$ 82.9 \pm 0.7 $ & 	$ 0.827 \pm 0.007 $ & 	$ 0.870 \pm 0.006 $ & 	$ 83.2 \pm 2.5 $ & 	$ 0.834 \pm 0.023 $ & 	$ 0.848 \pm 0.032 $ \\ 
\midrule
FitNet  \citep{DBLP:journals/corr/RomeroBKCGB14} & 	ImageNet & 	$ 83.1 \pm 0.7 $ & 	$ 0.826 \pm 0.009 $ & 	$ 0.886 \pm 0.005 $ & 	$ 83.2 \pm 4.1 $ & 	$ 0.836 \pm 0.038 $ & 	$ 0.853 \pm 0.035 $ \\
AT \citep{komodakis2017paying} & 	ImageNet & 	$ 83.6 \pm 0.7 $ & 	$ 0.832 \pm 0.009 $ & 	$ \textbf{0.889} \pm \textbf{0.002} $ & $ 86.0 \pm 1.6 $ & 	$ 0.861 \pm 0.014 $ & 	$ 0.882 \pm 0.028 $ \\ 
CC \citep{peng2019correlation} & 	ImageNet & 	$ 83.9 \pm 1.4 $ & 	$ 0.839 \pm 0.014 $ & 	$ 0.859 \pm 0.010 $ & 	$ 86.3 \pm 0.9 $ & 	$ 0.862 \pm 0.010 $ & 	$ \textbf{0.898} \pm \textbf{0.008} $ \\ 
CRD \citep{Tian2020Contrastive} & 	ImageNet & 	$ 81.6 \pm 3.0 $ & 	$ 0.808 \pm 0.038 $ & 	$ 0.885 \pm 0.008 $ & 	$ 84.9 \pm 2.7 $ & 	$ 0.846 \pm 0.030 $ & 	$ 0.890 \pm 0.017 $ \\ 
SemCKD  \citep{wang2022semckd} & 	ImageNet & 	$ 83.8 \pm 0.8 $ & 	$ 0.837 \pm 0.008 $ & 	$ 0.879 \pm 0.007 $ & 	$ 80.8 \pm 5.5 $ & 	$ 0.808 \pm 0.058 $ & 	$ 0.847 \pm 0.046 $ \\ 
MoMA (Ours) & 	ImageNet & 	$ \textbf{85.2} \pm \textbf{0.6} $ & 	$ \textbf{0.850} \pm \textbf{0.006} $ & 	$ 0.888 \pm 0.006 $ & 	$ \textbf{87.2} \pm \textbf{0.7} $ & 	$ \textbf{0.872} \pm \textbf{0.008} $ & 	$ 0.898 \pm 0.012 $ \\ 
MoMA (Ours) & 	PANDA & 	$ 82.5 \pm 0.5 $ & 	$ 0.822 \pm 0.004 $ & 	$ 0.867 \pm 0.006 $ & 	$ 84.2 \pm 0.9 $ & 	$ 0.844 \pm 0.008 $ & 	$ 0.861 \pm 0.020 $ \\ 

    \bottomrule
    \end{tabular}
    \end{adjustbox}
    \end{center}
	\end{table*}

         \begin{figure*}[t!]
        \centering
        \includegraphics[width=\textwidth]{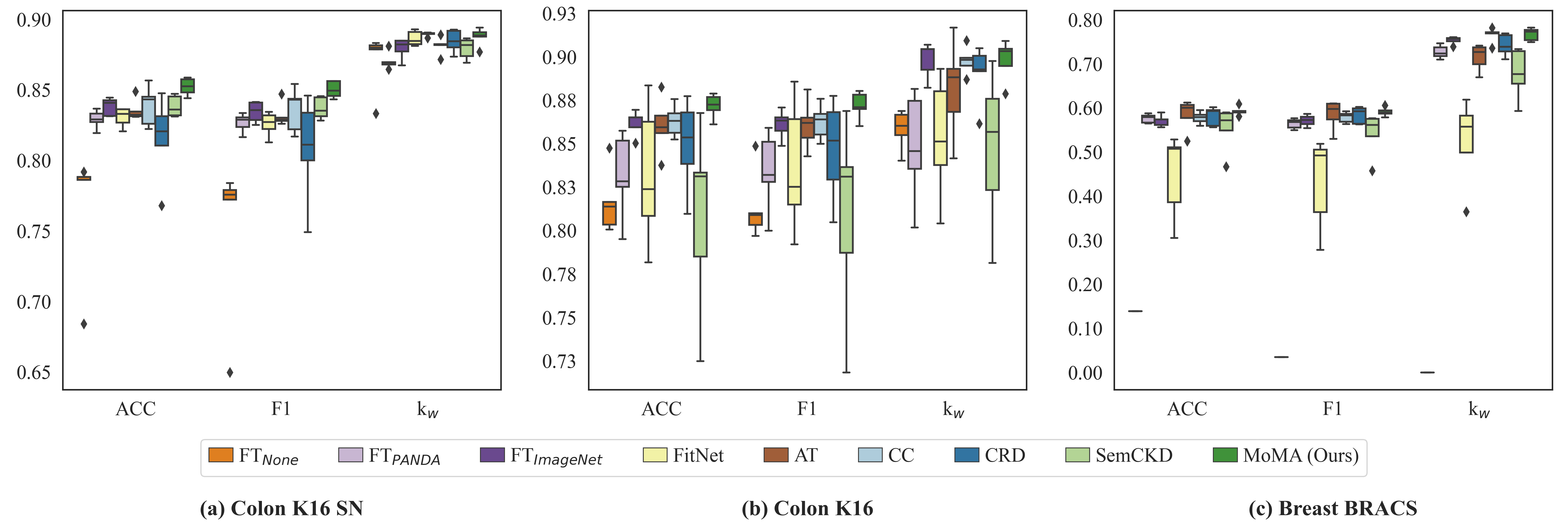}

        \caption{Bar plots for irrelevant task distillation. All the KD models utilize the pre-trained weights from ImageNet.}
        \label{fig:colon_boxplot}
    \end{figure*}

Table \ref{table:colon_result} and Fig. \ref{fig:colon_boxplot} (and Fig. \ref{fig:confusion_matric_number_colon_1} and Fig. \ref{fig:confusion_matric_number_colon_2} in the Appendix C) show the results of distillation from 4-class prostate cancer classification to colon tissue type classification.
Similar to the previous tasks, the student models either pre-trained on ImageNet FT$_{ImageNet}$ or PANDA FT$_{PANDA}$ were able to improve upon the model performance without pre-training. MoMA$_{ImageNet}$, utilizing the pre-trained weights of ImageNet, outperformed all the competing models except AT and CC in $\kappa_w$ for Colon K16 SN and Colon K16, respectively.
It is worth noting that, for both $TL$ and $FD$ approaches, the effect of the pre-trained weights of ImageNet was larger than that of PANDA. FT$_{ImageNet}$ was superior to FT$_{PANDA}$. Similarly, MoMA$_{ImageNet}$ obtained better performance than MoMA$_{PANDA}$. These results are

\subsection{Irrelevant task distillation: breast carcinoma sub-type classification}

    \begin{table*}[!t]
    \begin{center}
    \caption{Results of irrelevant task distillation: breast carcinoma sub-type classification.}
    \label{table:BRACS_result}
    \setlength{\tabcolsep}{6pt} % Default value: 6pt
	\renewcommand{\arraystretch}{1} % Default value: 1
	% \begin{adjustbox}{width=0.6\textwidth}
	\begin{tabular}{c|c|ccc}
% \cline{3-8}
     \toprule
Method &	Pretrained &	ACC ($\%$)  &	F1 &	$\kappa_w$ \\
\midrule
\citep{brancati2022bracs} &	     &	55.9 &	0.543 &	- \\
\midrule
FT &	None     &	$ 13.9 \pm 0.0 $ &	$ 0.035 \pm 0.000 $ &	$ 0.000 \pm 0.000 $ \\
FT &	ImageNet &	$ 57.1 \pm 1.3 $ &	$ 0.571 \pm 0.013 $ &	$ 0.753 \pm 0.009 $ \\
FT &	PANDA    &	$ 57.6 \pm 1.3 $ &	$ 0.564 \pm 0.011 $ &	$ 0.727 \pm 0.015 $ \\
\midrule
FitNet  \citep{DBLP:journals/corr/RomeroBKCGB14} & 	ImageNet &	$ 44.7 \pm 9.7 $ &	$ 0.431 \pm 0.106 $ &	$ 0.524 \pm 0.100 $ \\
AT \citep{komodakis2017paying} & 	ImageNet &	$ 58.4 \pm 3.6 $ &	$ 0.584 \pm 0.034 $ &	$ 0.715 \pm 0.030 $ \\
CC \citep{peng2019correlation} & 	ImageNet &	$ 57.4 \pm 1.4 $ &	$ 0.575 \pm 0.011 $ &	$ 0.748 \pm 0.014 $ \\
CRD \citep{Tian2020Contrastive} & 	ImageNet &	$ 58.1 \pm 2.1 $ &	$ 0.584 \pm 0.019 $ &	$ 0.742 \pm 0.025 $ \\
SemCKD  \citep{wang2022semckd} & 	ImageNet &	$ 55.3 \pm 5.1 $ &	$ 0.541 \pm 0.050 $ &	$ 0.677 \pm 0.058 $ \\
MoMA (Ours) & 	ImageNet &	$ \textbf{59.2} \pm \textbf{1.0} $ &	$ \textbf{0.591} \pm \textbf{0.010} $ &	$ \textbf{0.767} \pm \textbf{0.015} $ \\
MoMA (Ours) & 	PANDA    &	$ 57.9 \pm 0.6 $ &	$ 0.565 \pm 0.008 $ &	$ 0.718 \pm 0.025 $ \\

    \bottomrule
    \end{tabular}
    % \end{adjustbox}
    \end{center}
	\end{table*}

The results of distillation from 4-class prostate cancer classification to breast carcinoma sub-type classification are illustrated in Table \ref{table:BRACS_result} and Fig. \ref{fig:colon_boxplot} (and Fig. \ref{fig:confusion_matric_number_breast} in the Appendix C).
Unlike the distillation to colon tissue type classification, FT$_{PANDA}$ obtained slightly higher performance than FT$_{ImageNet}$, though the gap was minimal. When it comes to MoMA, the benefit of the pre-trained weighted on ImageNet was apparent. MoMA$_{ImageNet}$ substantially surpassed MoMA$_{PANDA}$ by 1.3 ACC, 0.026 F1, and 0.049 $\kappa_w$ on average. Moreover, other competitors were greatly inferior to MoMA$_{ImageNet}$, suggesting the utility of MoMA on the irrelevant task distillation.

\subsection{Inter- and intra-class correlations for student and teacher models}

    \begin{figure}[h]
        \centering
        \includegraphics[width=0.45\textwidth]{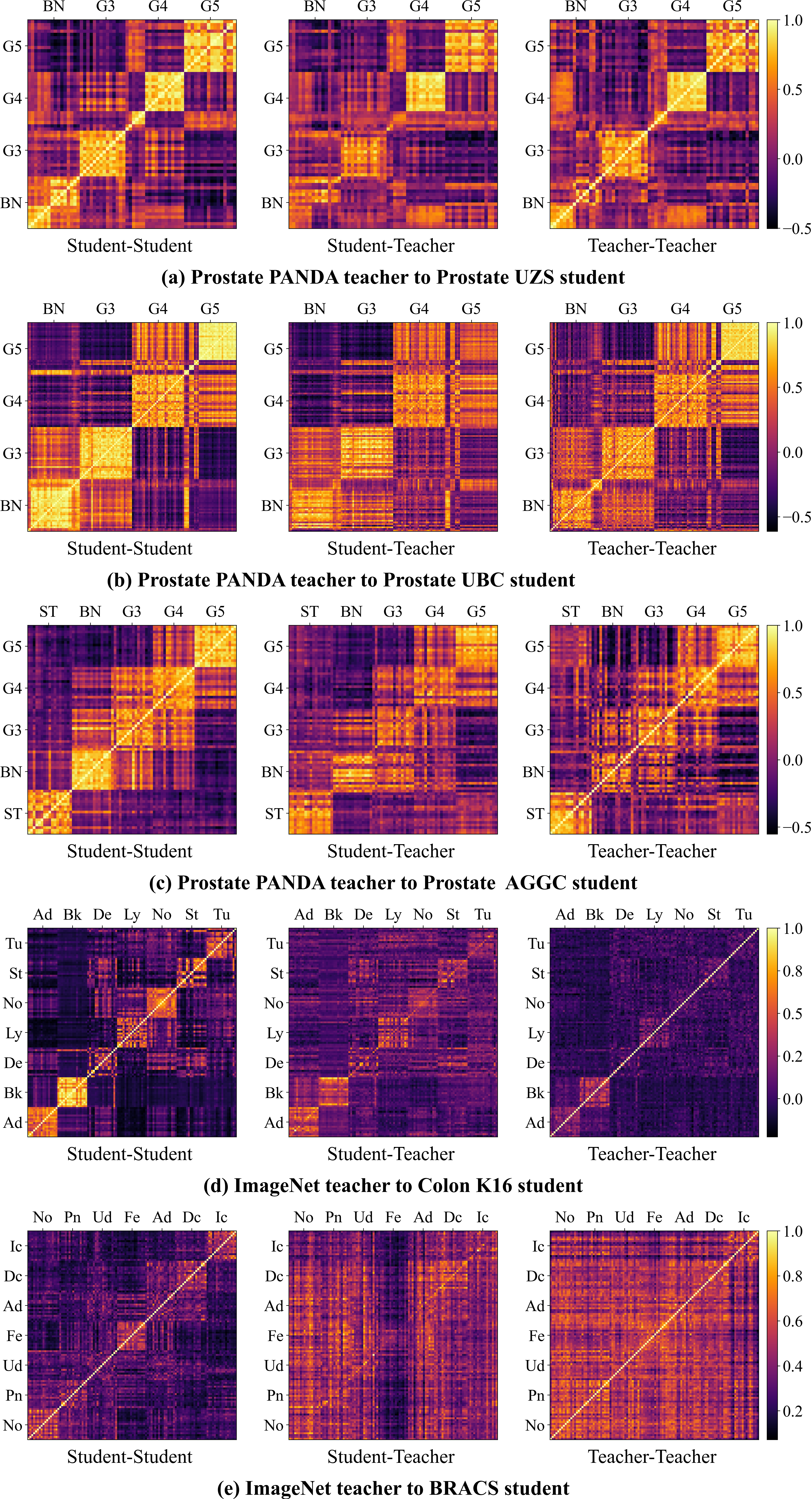}
        \caption{The correlation coefficient matrix between feature presentations of a teacher network and student network}
        \label{fig:correlation}
    \end{figure}

Fig. \ref{fig:correlation} shows the inter- and intra-class correlations between feature embeddings encoded by the MoMA student and teacher models. Three types of correlations were measured, including student-to-student, student-to-teacher, and teacher-to-teacher correlations. As for the teacher model, we chose the best teacher model per task, i.e., the teacher models pre-trained on PANDA for two prostate cancer classification tasks and the ImageNet teacher model for the colon tissue type classification task. For each task, 16 samples per class were randomly chosen from the validation set.

For the three distillation tasks, the student models, in general, showed higher intra-class correlations and lower inter-class correlations, which explains the superior performance of the student models in the classification tasks. For instance, in the same task distillation from PANDA to Prostate USZ, the PANDA teacher model was partially successful in demonstrating the connections between four different types of class labels; it had difficulties in distinguishing some samples in BN and G3, which can be shown by the lower intra-class correlations within BN and G3. This is likely due to variations between the source and target datasets. However, the student model, trained on a target/student dataset, was able to achieve stronger intra-class correlations for both BN and G3 while still maintaining high intra-class correlations for G4 and G5; inter-class correlations were lowered in general. 
In a head-to-head comparison between Prostate USZ and UBC, we found substantial differences between them. There were, in general, higher inter-class correlations on Prostate UBC, which partially explains the superior performance of TC$_{PANDA}$ on the dataset. The inter-class correlations were observed among most of the classes for Prostate USZ. On Prostate UBC, the inter-class correlations were found mostly between BN and G3 and between G4 and G5. 
As for other distillation tasks, we made similar observations. The intra-class correlations tend to improve as the model is trained on a student dataset.
Such improvement, achieved through the MoMA framework, is not only due to the knowledge from the teacher model but also due to the utilization of the target dataset.

\begin{figure*}[!t]
\centering
\includegraphics[height=0.95\textheight]{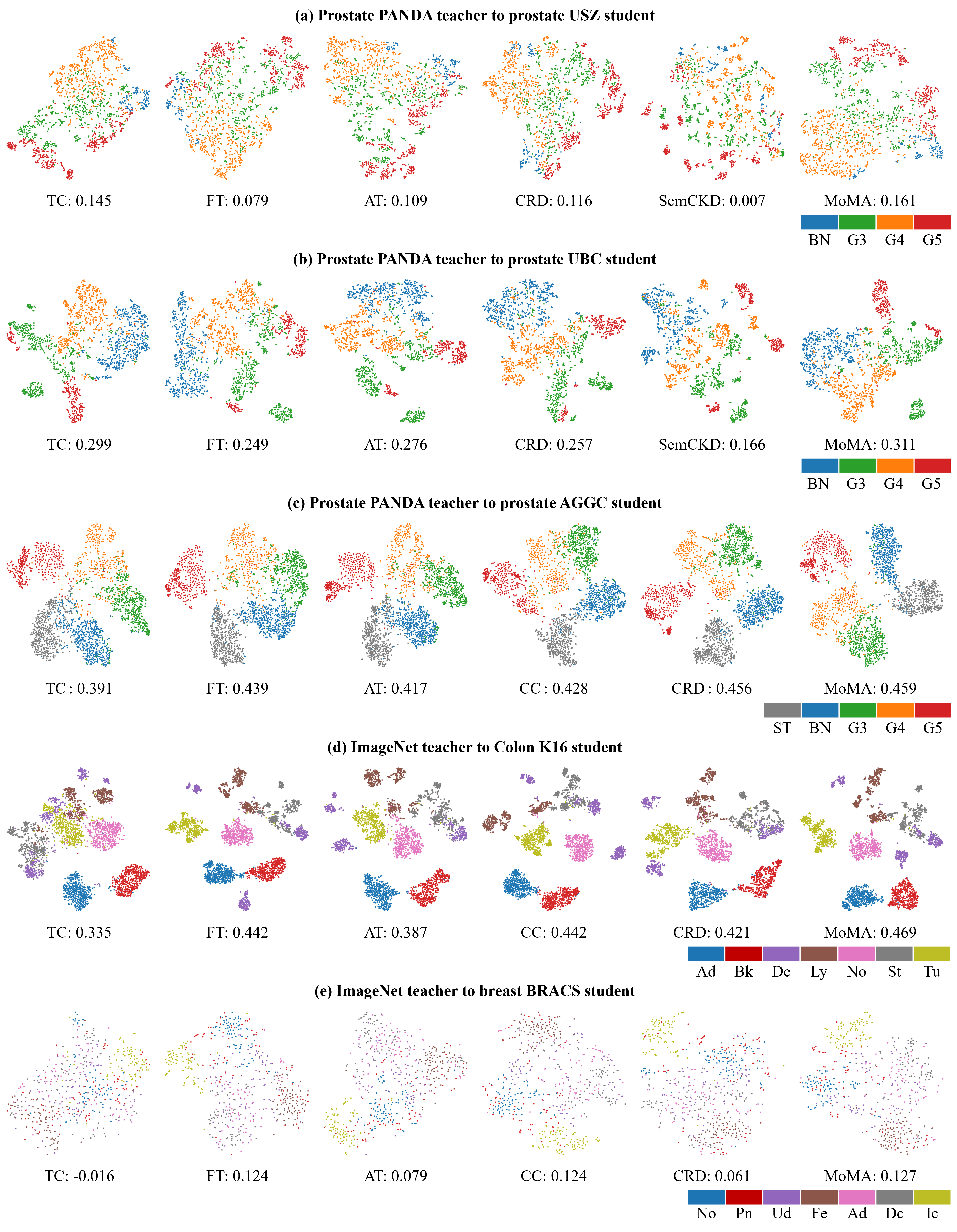}
\caption{t-SNE visualization of feature representations with silhouette scores for ImageNet and PANDA teacher models and 5 student datasets: (a) Prostate USZ, (b) Prostate UBC, (c) Prostate AGGC, (d) Colon K16, (e) Breast BRACS.}
\label{fig:tsne}
\end{figure*}

\subsection{Visualization of feature representation}

Fig. \ref{fig:tsne} exhibits the visualization and quantification of the feature representation of MoMA and other competing models. We have used t-SNE to visualize the feature representation and silhouette score to quantify it. Both visualization and quantification results consistently show that MoMA improves the feature representation in comparison to the teacher model and student models regardless of the training methods and datasets. However, the effect of other methods was inconsistent. No competing models were able to provide a constant improvement over the teacher model, highlighting the superior capability of MoMA in advancing the feature representation.

\subsection{Ablation study}

    \begin{table*}[!t]
    \begin{center}
    \caption{Ablation results of MoMA with and without Multi-head Attention on the three distillation tasks.}
    \label{table:ablation}
    \setlength{\tabcolsep}{6pt} % Default value: 6pt
	\renewcommand{\arraystretch}{1} % Default value: 1
	\begin{adjustbox}{width=\textwidth}
	\begin{tabular}{c|c|ccc|ccc}
\cline{3-8}
     % \toprule
\multicolumn{2}{c}{} & & Test I 	 & &	& Test II 	 &	 \\
\midrule
Task & Method &		ACC &	F1 &	$\kappa_w$ &	ACC &	F1 &	$\kappa_w$ \\
\midrule

Same task  & 	MoMA w/o MSA & 	$ 72.0 \pm 0.4 $ & 	$ 0.663 \pm 0.004 $ & 	$ 0.654 \pm 0.008 $ & 	$ 73.8 \pm 1.7 $ & 	$ 0.616 \pm 0.014 $ & 	$ \textbf{0.667} \pm \textbf{0.012} $ \\ 
distillation & 	MoMA & 	$ \textbf{73.6} \pm \textbf{1.0} $ & 	$\textbf{ 0.687} \pm \textbf{0.011} $ & 	$ \textbf{0.670} \pm \textbf{0.010 }$ & 	$ \textbf{75.5} \pm \textbf{1.8} $ & 	$ \textbf{0.622} \pm \textbf{0.019 }$ & 	$ 0.666 \pm 0.015 $ \\ 
\midrule
Relevant task  & 	MoMA w/o MSA & 	$ 75.6 \pm 1.2 $ & 	$ 0.649 \pm 0.036 $ & 	$ 0.786 \pm 0.045 $ & 	$ \textbf{77.2} \pm \textbf{2.4} $ & 	$ 0.601 \pm 0.020 $ & 	$ \textbf{ 0.804} \pm \textbf{0.042} $ \\ 
distillation & 	MoMA & 	$ \textbf{77.1} \pm\textbf{ 2.0} $ & 	$ \textbf{0.670} \pm \textbf{0.042} $ & 	$ \textbf{0.798} \pm \textbf{0.029} $ & 	$ 77.1 \pm 2.3 $ & 	$ \textbf{0.609} \pm \textbf{0.015} $ & 	$ 0.794 \pm 0.018 $ \\ 
\midrule
Irrelevant task  & 	MoMA w/o MSA & 	$ 84.0 \pm 0.4 $ & 	$ 0.838 \pm 0.004 $ & 	$ 0.880 \pm 0.005 $ & 	$ 85.5 \pm 1.4 $ & 	$ 0.856 \pm 0.015 $ & 	$ \textbf{0.900} \pm \textbf{0.008}$ \\ 
distillation: colon & 	MoMA & 	$ \textbf{85.2} \pm \textbf{0.6} $ & 	$ \textbf{0.850} \pm \textbf{0.006} $ & 	$ \textbf{0.888} \pm \textbf{0.006} $ & 	$ \textbf{87.2} \pm \textbf{0.7} $ & 	$\textbf{ 0.872} \pm \textbf{0.008} $ & 	$ 0.898 \pm  0.012 $ \\ 
 \cline{1-8}
Irrelevant task  & 	MoMA w/o MSA & 	$ 58.4 \pm 0.3 $ &	$ 0.585 \pm 0.003 $ &	$ 0.760 \pm 0.014 $   & 	 & 	 &  \\ 
distillation: breast & 	MoMA & 	$ \textbf{59.2} \pm \textbf{1.0} $ & 	$ \textbf{0.591} \pm \textbf{0.010} $ & 	$ \textbf{0.767} \pm \textbf{0.015} $ & 	 & 	 & 	 \\ 
\cline{1-5}
% Irrelevant task  & 	MoMA w/o MSA & $ 68.9 \pm 0.3 $ &	$ 0.629 \pm 0.003 $ &	$ 0.314 \pm 0.004 $ &
%  	 & 	 & 	 \\ 
% distillation: gastric & 	MoMA & $ \textbf{79.6} \pm \textbf{1.0} $ &	$ \textbf{0.746} \pm \textbf{0.010} $ &	$ \textbf{0.500} \pm \textbf{0.012} $ &
%  	 & 	 & 	\\ 
 \cline{1-5}
    \end{tabular}
    \end{adjustbox}
    \end{center}
\end{table*}

Table \ref{table:ablation} compares the performance of MoMA with and without MSA across three distillation tasks. The results demonstrate the crucial role of MSA in the proposed approach. 
MoMA without MSA, in general, experienced a performance drop for the three distillation tasks. Without MSA, MoMA was able to achieve better or comparable performance to other competing models across different distillation tasks, suggesting the effectiveness and robustness of the proposed framework.

Moreover, MoMA utilizes up to three loss functions, including $\mathcal{L}_{CE}$, $\mathcal{L}_{NCE}$, and $\mathcal{L}_{KL}$. The effect of these loss functions is already available in Table \ref{table:prostate_tma}, \ref{table:prostate_aggc}, \ref{table:colon_result}, and \ref{table:BRACS_result}. $\mathcal{L}_{KL}$ is only used for the same task distillation. Using $\mathcal{L}_{CE}$ only, the model becomes FT$_{PANDA}$ or FT$_{ImageNet}$ depending on the pre-trained weights. MoMA uses both $\mathcal{L}_{CE}$ and $\mathcal{L}_{NCE}$. KL+MoMA is the case where it exploits all three loss functions. We found that MoMA produces superior performance across the three distillation tasks, indicating the benefit of $\mathcal{L}_{NCE}$. The utility of $\mathcal{L}_{KL}$ can be found in the same task distillation.

\subsection{Effect of self-supervised learning}

\begin{table*}[!t]
    \begin{center}
    \caption{Results of three distillation tasks with MoMA and self-supervised learning.}
    \label{table:ssl_result}
    \setlength{\tabcolsep}{6pt} % Default value: 6pt
	\renewcommand{\arraystretch}{1} % Default value: 1
	\begin{adjustbox}{width=\textwidth}
	\begin{tabular}{c|c|ccc|ccc}
 \cline{3-8}
     %\toprule
     
\multicolumn{2}{c}{} & & Prostate USZ (Test I) & &	& Prostate UBC (Test II)	 &	 \\
\midrule
Method &	Pretrained &	ACC ($\%$)  &	F1 &	$\kappa_w$ &	ACC ($\%$)  &	F1 &	$\kappa_w$ \\
\midrule
%\multicolumn{2}{c}{} & & Prostate USZ (Test I) & &	& Prostate UBC (Test II)	 &	 \\
%\midrule
FT &	CTransPath &	$ 70.6 \pm 3.2 $ &	$ 0.635 \pm 0.037 $ &	$ 0.618 \pm 0.035 $ &	$ 76.4 \pm 3.2 $ &	$ 0.657 \pm 0.037 $ &	$ 0.700 \pm 0.035 $ \\
MoMA &	CTransPath &	$ \textbf{71.6} \pm \textbf{1.5} $ &	$ \textbf{0.652} \pm \textbf{0.019} $ &	$ \textbf{0.633} \pm \textbf{0.02} $ &	$ \textbf{78.2} \pm \textbf{1.3} $ &	$ \textbf{0.682} \pm \textbf{0.025} $ &	$ \textbf{0.718} \pm \textbf{0.014} $ \\
\midrule
FT &	Lunit$_{DINO}$ &	$ 69.7 \pm 0.6 $ &	$ 0.629 \pm 0.007 $ &	$ 0.618 \pm 0.008 $ &	$ 75.7 \pm 2.2 $ &	$ 0.649 \pm 0.017 $ &	$ 0.690 \pm 0.030 $ \\
MoMA &	Lunit$_{DINO}$ &	$ \textbf{70.8} \pm \textbf{1.4} $ &	$ \textbf{0.638} \pm \textbf{0.012} $ &	$ \textbf{0.622} \pm \textbf{0.012} $ &	$ \textbf{76.9} \pm \textbf{02.5} $ &	$ \textbf{0.667} \pm \textbf{0.035} $ &	$ \textbf{0.702} \pm \textbf{0.019} $ \\

\midrule
\multicolumn{2}{c}{} & &  AGGC CV (Test I)	 &  &	& AGGC test (Test II)	 & 	 \\
\midrule
Method &	Pretrained &	ACC ($\%$)  &	F1 &	$\kappa_w$ &	ACC ($\%$)  &	F1 &	$\kappa_w$ \\
\midrule
FT &	CTransPath &	$ \textbf{76.3} \pm \textbf{3.6} $ &	$ \textbf{0.648} \pm \textbf{0.069} $ &	$ \textbf{0.805} \pm \textbf{0.032} $ &	$ \textbf{77.9} \pm \textbf{1.9} $ &	$ 0.604 \pm 0.012 $ &	$ 0.793 \pm 0.048 $ \\
MoMA &	CTransPath &	$75.8 \pm 4.4 $ &	$ 0.647 \pm 0.066 $ &	$ \textbf{0.805} \pm \textbf{0.032} $ &	$ 77.8 \pm 2.6 $ &	$ \textbf{0.605} \pm \textbf{0.015} $ &	$ \textbf{0.807} \pm \textbf{0.051} $ \\
\midrule
FT &	Lunit$_{DINO}$ &	$ 71.1 \pm 4.3 $ &	$ 0.595 \pm 0.066 $ &	$ 0.743 \pm 0.028 $ &	$ 63.3 \pm 06.9 $ &	$ 0.493 \pm 0.040 $ &	$ 0.482 \pm 0.190 $ \\
MoMA &	Lunit$_{DINO}$ &	$ \textbf{73.8} \pm \textbf{4.1} $ &	$ \textbf{0.636} \pm \textbf{0.075} $ &	$ \textbf{0.781} \pm \textbf{0.033} $ &	$ \textbf{73.1} \pm \textbf{3.0} $ &	$ \textbf{0.572} \pm \textbf{0.012} $ &	$ \textbf{0.777} \pm \textbf{0.05} $ \\
\midrule
\multicolumn{2}{c}{} & & Colon K16 SN (Test I) & &	& Colon K16 (Test II) 	 & \\
\midrule
Method &	Pretrained &	ACC ($\%$)  &	F1 &	$\kappa_w$ &	ACC ($\%$)  &	F1 &	$\kappa_w$ \\
\midrule
FT &	CTransPath &	$ 71.6 \pm 1.5 $ &	$ 0.581 \pm 0.012 $ &	$ 0.842 \pm 0.007 $ &	$ 74.2 \pm 1.4 $ &	$ 0.609 \pm 0.010 $ &	$ 0.878 \pm 0.013 $  \\
MoMA &	CTransPath &	$ \textbf{83.6} \pm \textbf{2.5} $ &	$ \textbf{0.832} \pm \textbf{0.028} $ &	$ \textbf{0.899} \pm \textbf{0.012} $ &	$ \textbf{89.1} \pm \textbf{1.9} $ &	$ \textbf{0.890} \pm \textbf{0.022} $ &	$ \textbf{0.927} \pm \textbf{0.007} $\\
\midrule
FT &	Lunit$_{DINO}$ &	$ \textbf{83.9} \pm \textbf{2.0} $ &	$ \textbf{0.836} \pm \textbf{0.020} $ &	$ \textbf{0.867} \pm \textbf{0.018} $ &	$ 72.5 \pm 1.7 $ &	$ 0.597 \pm 0.011 $ &	$ 0.859 \pm 0.008 $ \\
MoMA &	Lunit$_{DINO}$ &	$ 82.1 \pm 1.8 $ &	$ 0.817 \pm 0.021 $ &	$ 0.852 \pm 0.013 $ &	$ \textbf{88.4} \pm \textbf{1.4} $ &	$ \textbf{0.884} \pm \textbf{0.014} $ &	$ \textbf{0.918} \pm \textbf{0.003} $ \\
\midrule
\multicolumn{2}{c}{} & & Breast BRACS & &	& 	 \\

 \cline{1-5}
Method &	Pretrained &	ACC ($\%$)  &	F1 &	$\kappa_w$ &	  &	 &	 \\
 \cline{1-5}
FT &	CTransPath &	 $ 63.5 \pm 1.3 $ &	$ 0.628 \pm 0.016 $ &	$ \textbf{0.818} \pm \textbf{0.011} $  &&&\\
MoMA &	CTransPath &	 $ \textbf{63.9} \pm \textbf{1.3} $ &	$ \textbf{0.633} \pm \textbf{0.015} $ &	$ 0.807 \pm 0.014 $ &&&\\

 \cline{1-5}
FT &	Lunit$_{DINO}$ & $ 62.3 \pm 0.7 $ &	$ 0.622 \pm 0.007 $ &	$ \textbf{0.826} \pm \textbf{0.011} $\\
MoMA &	Lunit$_{DINO}$ & $ \textbf{63.2} \pm \textbf{1.3} $ &	$ \textbf{0.630} \pm \textbf{0.013} $ &	$ 0.806 \pm 0.013 $ &&& \\

 \cline{1-5}
\end{tabular}
\end{adjustbox}
\end{center}
\end{table*}

\begin{table*}[!t]
    \begin{center}
    \caption{Results of irrelevant task distillation: gastric microsatellite instability prediction.}
    \label{table:result_stad}
    \setlength{\tabcolsep}{6pt} % Default value: 6pt
	\renewcommand{\arraystretch}{1} % Default value: 1
	% \begin{adjustbox}{width=0.6\textwidth}
	\begin{tabular}{c|c|ccc}
% \cline{3-8}
     \toprule
Method &	Pretrained &	ACC ($\%$)  &	F1 &	$\kappa_w$ \\
\midrule
FT &	CTransPath &	$ 65.2 \pm 23.2 $ &	$ 0.585 \pm 0.225 $ &	$ 0.296 \pm 0.199 $ \\
\midrule
AT \citep{komodakis2017paying} & 	CTransPath &	$ 77.9 \pm 4.2 $ &	$ 0.718 \pm 0.051 $ &	$ 0.438 \pm 0.101 $ \\
CC \citep{peng2019correlation} & 	CTransPath &	$ 76.7 \pm 2.5 $ &	$ 0.693 \pm 0.026 $ &	$ 0.389 \pm 0.054 $ \\
CRD \citep{Tian2020Contrastive} & 	CTransPath &	$ 75.1 \pm 3.0 $ &	$ 0.633 \pm 0.113 $ &	$ 0.301 \pm 0.169 $ \\
MoMA (Ours) & 	CTransPath &	$ \textbf{79.6} \pm \textbf{2.7} $ &	$ \textbf{0.746} \pm \textbf{0.012} $ &	$ \textbf{0.500} \pm \textbf{0.018} $ \\

    \bottomrule
    \end{tabular}
    % \end{adjustbox}
    \end{center}
\end{table*}

In order to further demonstrate the utility of MoMA, we employed two Transformer-based models, i.e., CTransPath \citep{wang2022transformer} and Lunit$_{DINO}$ \citep{kang2023benchmarking}, that were trained on a large pathology image data via self-supervised learning. CTransPath is a hybrid CNN-transformer backbone that replaces the patch partition part with a CNN module in a tiny Swin Transformer \citep{liu2021swin} model. CTransPath was trained on around 15 million unlabeled patches cropped from WSIs in TCGA and PAIP. Lunit$_{DINO}$ is a small ViT network that was trained on a total of 32.6 million image patches via DINO \citep{caron2021emerging}.
Employing these two models, we conducted the three distillation tasks with and without MoMA. The results are shown in Table \ref{table:ssl_result}. In a head-to-head comparison, the two models with MoMA, by and large, obtained superior performance in comparison to the models without MoMA. Specifically, CTransPath with MoMA substantially outperformed CTransPath without MoMA (FT) on the same task distillation and irrelevant task distillation. On the relevant task distillation, CTransPath with MoMA was slightly inferior to CTransPath without MoMA such as, on average, -0.005 ACC and -0.001 F1 for Prostate AGGC CV and -0.001 ACC for Prostate AGGC test. 
Lunit$_{DINO}$ with MoMA also surpassed Lunit$_{DINO}$ without MoMA for the three distillation tasks except Colon K16 SN. Comparing CTransPath and Lunit$_{DINO}$, CTransPath almost always produced better results than Lunit$_{DINO}$. Overall, these results suggest that MoMA is generic and thus can be applied to various tasks and problems regardless of architectures and pre-training methodologies, highlighting the utility and usefulness of MoMA in learning a target model.

Moreover, we compared the pre-trained weights from CTransPath, and Lunit$_{DINO}$ against those from ImageNet and PANDA in regard to the classification performance with MoMA (Table \ref{table:prostate_tma}, \ref{table:prostate_aggc}, and  \ref{table:ssl_result}). Pre-trained weights from ImageNet and PANDA were obtained using a CNN (EfficnetNet-b0) via supervised learning, and those from CTransPath and Lunit$_{DINO}$ were acquired using Transformer models via self-supervised learning. 
For the same task distillation, MoMA$_{CTransPath}$ and MoMA$_{Lunit_{DINO}}$ were shown to be inferior to MoMA$_{PANDA}$; on Prostate USZ, both models substantially underperformed MoMA$_{PANDA}$ by $\le$0.020 ACC, $\le$0.035, and $\le$0.047 on average; on Prostaet UBC, they were superior to MoMA$_{PANDA}$ but were inferior to KL+MoMA$_{PANDA}$ by ACC and $\kappa_w$.
As for the relevant distillation task, MoMA$_{PANDA}$ was better or comparable to MoMA$_{CTransPath}$ and MoMA$_{Lunit_{DINO}}$. MoMA$_{Lunit_{DINO}}$ was always inferior to MoMA$_{PANDA}$. MoMA$_{CTransPath}$ outperformed MoMA$_{PANDA}$ on Prostate AGGC CV by $\kappa_w$ and on Prostate AGGC test by ACC and $\kappa_w$. 
For the irrelevant distillation task, CTransPath and Lunit$_{DINO}$ were superior to ImageNet and PANDA. On Colon K16 SN, MoMA$_{ImageNet}$ outperformed both MoMA$_{CTransPath}$ and MoMA$_{Lunit_{DINO}}$ for all evaluation metrics except $\kappa_w$ by MoMA$_{CTransPath}$; however, both MoMA$_{CTransPath}$ and MoMA$_{Lunit_{DINO}}$ substantially surpassed MoMA$_{ImageNet}$ and MoMA$_{PANDA}$, on average, by $\le$1.9\% ACC, $\le$0.018 F1, and $\le$0.029 $\kappa_w$ on Colon K16 and by $\le$4.7\% ACC, $\le$0.042 F1, and $\le$0.040 $\kappa_w$ on BRACS. These results suggest that the same task distillation benefits from the same teacher dataset and the irrelevant task distillation benefits from the more general teacher dataset (ImageNet, CTransPath, and Lunit$_{DINO}$).

Furthermore, we utilized CTransPath to conduct the WSI-level MSI classification, which is another irrelevant task distillation, due to its excellent performance on the irrelevant task distillation. 
Table. \ref{table:result_stad} (and Fig. \ref{fig:confusion_matric_number_gastric} in the Appendix C) demonstrates the results of the MSI classification.
MoMA$_{CTransPath}$ achieved an ACC of 79.6, F1 of 0.746, and $\kappa_w$ of 0.500, on average, which is superior to other competing models by $\ge$1.7\% ACC, $\ge$0.028 F1, and $\ge$0.062 $\kappa_w$ on average. AT$_{CTransPath}$ was shown to be the second-best model over all three evaluation metrics. The fine-tuning approach (FT$_{CTransPath}$) was greatly worse than all the KD approaches.

\section{Discussion} \label{section:discussion}
In this work, we introduce an approach of KD, so-called MoMA, to build an optimal model for a target histopathology dataset using the existing models on similar, relevant, and irrelevant tasks. The experiment results show that MoMA offers advancements in distilling knowledge for the three classification tasks. Regardless of the type of the distillation tasks, MoMA enables the student model to inherit feature extraction capabilities from the teacher model and to conduct accurate classification for the target task. Exploiting the knowledge from both the source and target datasets, MoMA also provides superior generalizability on unseen, independent test sets across three different tasks.

We conduct KD from a range of teacher models, including TC$_{PANDA}$ (supervised pathology-domain-specific teacher model), TC$_{CTransPath}$ and TC$_{Lunit{DINO}}$ (self-supervised pathology-general teacher model), and TC$_{ImageNet}$ (supervised general teacher model), to address four domain-specific tasks. While the concept of task-specific distillation \citep{jiao2019tinybert, kim2020fastformers, tang2019distilling, sanh2019distilbert} is prevalent in natural language processing, its application in computer vision and computational pathology is less common. One notable example in the field of NLP is TinyBERT \citep{jiao2019tinybert}, which employs task-specific distillation both during the pre-training and fine-tuning stages, augmenting data and distilling knowledge from a fine-tuned BERT teacher model. Additionally, recent work by \citep{quteineh2022enhancing} introduces language generation techniques to enhance task-specific distillation, leveraging synthesized samples to improve student performance.

Our proposed MoMA framework offers a flexible approach to knowledge distillation, capable of performing general distillation with or without task-specific teacher models. Specifically, in the scenario where TC$_{PANDA}$ distills knowledge to other prostate models, our method parallels the task-specific distillation process seen in TinyBERT. However, our framework demonstrates efficacy even without task-specific teacher models such as TC$_{ImageNet}$, TC$_{CTransPath}$, or TC$_{Lunit_{DINO}}$. Furthermore, our experimental results align with findings in task-specific distillation literature, showing that a specific teacher likes TC$_{PANDA}$ yields superior performance compared to a general teacher like TC$_{ImageNet}$ when applied to the same task or related tasks. This comparison highlights the versatility and effectiveness of our proposed approach in leveraging knowledge from diverse teacher models.

The aim of this work is to propose a method to distill the knowledge from the source/teacher domain to the target domain without direct access to the source data. Transferring only the teacher model is more feasible in various contexts; for instance, when the source data is enormous in size, like ImageNet, it is time-consuming and inefficient to train a model using such source data; healthcare data, including pathology images, is restricted for security and privacy reasons, and thus transferring to target data centers or hospitals is likely to be infeasible. 
In such circumstances, KD is a key to resolve all the related issues. 
In the distillation, we emphasize that the choice of the pre-trained teacher model is crucial as it directly impacts the performance of the target model. 
Based on the experimental results across different classification tasks, it is evident that the better the teacher model is, the greater benefits to the student model it provides.

The proposed MoMA framework is an end-to-end training approach, eliminating the need for extensive training of the self-supervised task followed by fine-tuning on labeled datasets. Moreover, self-supervised methods require a large amount of training data which is not always available in medical/pathology image analysis. Leveraging the high-quality teacher model through MoMA facilitates robust training and convergence of the student model on smaller target datasets. Furthermore, the excellent performance in the relevant and irrelevant tasks suggests that MoMA could be utilized solely as a feature-embedding distillation mechanism without requiring a meticulous redesign of the model architecture and distillation framework in response to the specific requirement of downstream tasks. 

In the ablation study, the role of MSA was apparent in MoMA. The previous SSCL assumed that all samples are equally important. However, depending on the appearance and characteristics of an image and the extent of augmentation applied, the classification task may become easier or more challenging. By incorporating MSA, MoMA gains the ability to selectively focus on important samples while allocating less attention to other samples. It is worth noting that the contrastive loss $\mathcal{L}_{NCE}$ does not treat each input sample independently, unlike the supervised cross-entropy loss $\mathcal{L}_{CE}$. With MSA, MoMA learns the relationships among samples within a batch before they are fed into the self-supervised contrastive loss, and, in turn, used to update the queue $Z^{queue}$, prioritizing and enriching the information within these samples and allowing for more effective optimization of the model.

MoMA was not able to obtain the best performance for all the datasets and evaluation metrics (Table \ref{table:prostate_tma}, \ref{table:prostate_aggc}, and \ref{table:colon_result}). However, none of the competing methods were as good as MoMA across the three distillation tasks. The performance of other competitors varied greatly across the datasets. Excluding MoMA, the best model for each dataset was different from one to the other. For example, in Prostate USZ, FT$_{PANDA}$ was the best model using ACC, F1, and $\kappa_w$; in Prostate UBC, Vanilla KD (ACC) and KL+SemCKD (F1 and $\kappa_w$) were the best models; in AGGC CV, CC (ACC and F1$_w$) and CRD ($\kappa_w$) were the best models; in AGGC test, CRD (ACC and $\kappa_w$) and FT$_{PANDA}$ (F1$_w$) were the best models; in Colon K16 SN, CC (ACC and $\kappa_w$) and AT (F1) were the best models; in Colon K16, CC was the best model. Such inconsistency in the performance by the competitors demonstrates the level of difficulty of the classification tasks in this study, involving a number of independent datasets obtained from differing sources and/or imaging devices. This further emphasizes the capability of MoMA to achieve the best or comparable results across multiple tasks and datasets with sufficient variations.

In a close examination of the classification results by MoMA, we found that the classification results vary across the datasets and these may be ascribable to data/class imbalance. For the same task distillation, in particular, the data/class distribution greatly differs between the source (PANDA) and target datasets (Prostate USZ and UBC). In PANDA, G4 accounts for $\sim$50\% for both training and test sets. In Prostate USZ, G4 takes up $<$30\% for the training and validation sets but $\sim$50\% for the test set. It is $\sim$57\% for Prostate UBC. The ratio of other classes substantially differs from one to the other. Such differences may contribute to the difference in the overall classification performance. This may also affect the impact of $\mathcal{L}_{KL}$ in the classification performance. Since $\mathcal{L}_{KL}$ encourages to directly match the output logits of the teacher and student models, models with $\mathcal{L}_{KL}$ learn more from PANDA, leading to improved performance in Prostate UBC. 
For instance, KL+FitNet$_{PANDA}$, KL+CC$_{PANDA}$, KL+CRD$_{PANDA}$, KL+SemCKD$_{PANDA}$, KL+MoMA$_{ImageNet}$, and KL+MoMA$_{PANDA}$ substantially enhance performance for G4 compared to their respective counterparts (Fig. \ref{fig:confusion_matric_number_prostate_2} in the Appendix C). 
However, due to the difference between PANDA and Prostate USZ, the more the models learn from PANDA, the worse performance they obtain on Prostate USZ. This explains the comparable performance of FT$_{PANDA}$ on Prostate USZ but worse performance on Prostate UBC in comparison to that of MoMA. 
Despite the effects of data/class imbalance and loss functions, MoMA exhibited a substantial performance disparity between Prostate USZ and Prostate UBC, highlighting a limitation of the method.

As for the relevant and irrelevant tasks, not only the characteristics of the class labels but also the number of classes changes such as 4 classes in the same task distillation, 5 classes in the relevant task distillation, and 9 classes in the irrelevant task distillation. The superior results on these tasks and datasets suggest the strength and robustness of MoMA to the imbalance and disproportion in the data/class. 

The results of the three distillation tasks, i.e., the same task, relevant task, and irrelevant task, provide insights into the distillation of knowledge from the source domain to the target domain and the model development for the target domain. 
First, supervised learning on the target domain provides comparable performance in all three distillation tasks, but its performance on unseen data, i.e., generalizability, is not guaranteed. 
Second, the usage of the pre-trained weights is crucial for both TL and KD, regardless of the type of distillation tasks. 
Third, the effect of the pre-trained weights depends on the type of distillation tasks. As for the same and relevant tasks, the pre-trained weights from the same or relevant tasks were more useful. For the irrelevant task, the pre-trained weights from ImageNet were more beneficial than those from PANDA. Similarly, the pre-trained weights from Colon K19 were not useful for prostate cancer classification, which is shown in the Appendix B (Table B3). These indicate that not all pathology image datasets will be helpful in building a model for a specific computational pathology task and a dataset. 
Last, the KD strategy varies across different distillation tasks. The same task distillation takes advantage of the logits distillation, the relevant task distillation exploits the pre-trained weights, and the irrelevant task distillation does not make use of the (irrelevant) domain-specific knowledge much.

There are several limitations in our work. 
First, MoMA is designed to distill knowledge from a single source/teacher model. Several teacher models are available these days. The more teacher models we use, the better knowledge the student models may obtain. The follow-up study will extend MoMA to distill knowledge from multiple teacher models to train a single target/student model.
Second, four pathology image classification tasks including four organs were considered in this study. The effect of KD may vary depending on the type of tasks and organs.
Third, there exist other types of image classification tasks in computational pathology such as survival/outcome prediction. In general, the amount of survival/outcome dataset is smaller than that of cancer and tissue classifications, and thus KD may play a crucial role in survival/outcome prediction. 
Fourth, we adopt EfficientNet-B0 as the baseline model, which has been successfully adopted for many applications in computational pathology \citep{laleh2022benchmarking, chhipa2023magnification, jahanifar2021stain}.  Though several pathology-specific models have been proposed and shown to be effective in improving diagnostic performance \citep{koohbanani2021self, kong2022efficient, nair2022graph, le2021joint}, the primary focus of this study is to investigate the effect of the knowledge distillation framework. The effect of the proposed framework on other pathology-specific models needs to be further explored. 
Last, we only consider the same architecture distillation for the three distillation tasks.  Evaluating the proposed MoMA on different teacher-student combinations like ViT teacher to ViT student and CNN teacher to ViT student could provide valuable insights into the effectiveness of the proposed method across different architectures. In order to focus on KD, we conduct our study on the same architecture distillation and leave the study involving various architectures distillation for future research.

% Our knowledge distillation requires us to train the entire student network; meanwhile, other recent methods such as Low-Rank Adaptation (LORA) only require training a smaller number of parameters at injects trainable rank decomposition matrices or Retrieval-Augmented Generation (RAG) 

\section{Conclusions} \label{section:conclusion}
Herein, we propose an efficient and effective learning framework called MoMA to build an accurate and robust classification model in pathology images. Exploiting the KD framework, momentum contrastive learning, and SA, MoMA was able to transfer knowledge from a source domain to a target domain and to learn a robust classification model for five different tasks. Moreover, the experimental results of MoMA suggest an adequate learning strategy for different distillation tasks and scenarios. We anticipate that this will be a great help in developing computational pathology tools for various tasks. Future studies will entail the further investigation of the efficient KD method and extended validation and application of MoMA to other types of datasets and tasks in computational pathology.

\section*{Acknowledgments}
This work was supported by the National Research Foundation of Korea (NRF) (No. 2021R1A2C2014557) and by the Ministry of Trade, Industry and Energy (MOTIE) and Korea Institute for Advancement of Technology (KIAT) through the International Cooperative R\&D program (No. P0022543).

\bibliographystyle{IEEEtran}
\bibliography{ms}

\setcounter{table}{0}
\renewcommand{\thetable}{A\arabic{table}}
\appendix
\section*{Appendix A. Pseudocode of MoMA in a PyTorch-like style} 
\begin{algorithm}[h]
\scriptsize
\caption{Pseudocode of MoMA in a PyTorch-like style.}
\SetAlgoLined
    \PyComment{f\_S, f\_T: student and teacher encoder networks} \\
    \PyComment{z\_queue: a memory bank as a queue of Q feature vectors of size D (DxQ)} \\
    \PyComment{m: momentum, t: temperature} \\
    % \PyComment{t: temperature} \\

    f\_T.load(teacher\_weight) \PyComment{load teacher weight} \\
    f\_S.load(pretrained\_weight) \PyComment{load student pre-trained weight} \\
    \PyCode{for (x, label) in loader:} \PyComment{load a mini-batch x with N samples} \\
        \Indp   % start indent
        z\_S, logit\_S = f\_S.forward(x) \PyComment{ student feature vectors (NxD)} \\
        z\_T, logit\_T = f\_T.forward(x) \PyComment{ teacher feature vectors (NxD)} \\
        z\_T = z\_T.detach() \PyComment{no gradient to teacher feature vectors} \\
        
        \PyComment{gather teacher feature vectors from all GPUs} \\ 
        all\_z\_T = gather(z\_T) if num(GPUs) > 1 else z\_T \\

        \PyComment{instance level self-attention} \\
        z\_S = self\_att\_S(z\_S) \\
        z\_T = self\_att\_T(z\_T) \\
        all\_z\_T = self\_att\_all(all\_z\_T) \\

        \PyComment{contrastive loss, Eqn.(1)} \\
        loss\_NCE = NCELoss(z\_S, z\_T, z\_queue, t) \\
        \PyComment{classification loss} \\
        loss\_CE = CELoss(logit\_S, label) \\
        \PyComment{SGD update: student network} \\
        loss = loss\_NCE + loss\_CE \\
        loss.backward() \\
        update(f\_S.params) \\

        \PyComment{momentum update: teacher network} \\
        f\_T.params = m*f\_T.params + (1-m)*f\_S.params
         \texttt{\\} % insert a space
        
        \PyComment{update dictionary} \\
        enqueue(z\_queue, all\_z\_T) \PyComment{enqueue the current mini-batch} \\
        dequeue(z\_queue) \PyComment{ dequeue the earliest mini-batch} \\
        \Indm % end indent, must end with this, else all the below text will be indented
%    \PyCodeBluee{The differences between MoMA-KD and MoCo are highlighted in blue.}
\label{algo:algo}

\end{algorithm}
% \small{The differences between MoMA-KD and MoCo are highlighted in blue.}

\section*{Appendix B. Results of same task distillation  with different sizes of training data.} 

\setcounter{table}{0}
\renewcommand{\thetable}{B\arabic{table}}
    \begin{table*}[t!]
    \begin{center}
    \caption{Results of same task distillation on 50\% training data. KL denotes the use of KL divergence loss.}
    \label{table:prostate_tma_50}
    \setlength{\tabcolsep}{6pt} % Default value: 6pt
	\renewcommand{\arraystretch}{1} % Default value: 1
	\begin{adjustbox}{width=\textwidth}
	\begin{tabular}{c|c|ccc|ccc}
% \cline{3-8}
\cline{3-8}
 
\multicolumn{2}{c}{} & & Prostate USZ (Test I) & &	& Prostate UBC (Test II)	 &	 \\
\toprule
Method &	Pretrained &	ACC($\%$) &	F1 &	$\kappa_w$ &	ACC($\%$) &	F1 &	$\kappa_w$ \\
     \toprule

TC$_{PANDA}$&	ImageNet  &	$ 63.4  $ &	$ 0.526  $ &	$ 0.531  $ &	$ 78.2 $ &	$ 0.58 0 $ &	$ 0.680 $ \\
FT          &	None      &	$ 64.7 \pm 1.1 $ &	$ 0.543 \pm 0.026 $ &	$ 0.521 \pm 0.034 $ &	$ 16.5 \pm 3.6 $ &	$ 0.123 \pm 0.027 $ &	$ 0.058 \pm 0.035 $ \\
FT          &	ImageNet  &	$ 68.8 \pm 1.7 $ &	$ 0.617 \pm 0.013 $ &	$ 0.609 \pm 0.023 $ &	$ 70.3 \pm 4.3 $ &	$ 0.576 \pm 0.042 $ &	$ 0.585 \pm 0.086 $ \\
FT &	PANDA             &	$ 71.8 \pm 0.3 $ &	$ \textbf{0.665} \pm \textbf{0.012} $ &	$ 0.660 \pm 0.008 $ &	$ 68.6 \pm 4.3 $ &	$ 0.557 \pm 0.037 $ &	$ 0.591 \pm 0.061 $ \\
\midrule
FitNet \citep{DBLP:journals/corr/RomeroBKCGB14}  &	PANDA &	$ 63.9 \pm 4.2 $ &	$ 0.570 \pm 0.050 $ &	$ 0.560 \pm 0.065 $ &	$ 26.6 \pm 17.5 $ &	$ 0.197 \pm 0.143 $ &	$ 0.107 \pm 0.103 $ \\
AT \citep{komodakis2017paying}  &	PANDA &	$ 70.6 \pm 0.6 $ &	$ 0.635 \pm 0.013 $ &	$ 0.629 \pm 0.018 $ &	$ 76.9 \pm 2.0 $ &	$ 0.640 \pm 0.029 $ &	$ 0.671 \pm 0.011 $ \\
CC \citep{peng2019correlation} &	PANDA         &	$ 71.0 \pm 1.9 $ &	$ 0.660 \pm 0.012 $ &	$ 0.654 \pm 0.010 $ &	$ 74.0 \pm 2.7 $ &	$ 0.604 \pm 0.034 $ &	$ 0.632 \pm 0.074 $ \\
CRD \citep{Tian2020Contrastive} &	PANDA     &	$ 71.0 \pm 1.5 $ &	$ 0.644 \pm 0.027 $ &	$ 0.626 \pm 0.040 $ &	$ 70.4 \pm 6.0 $ &	$ 0.567 \pm 0.066 $ &	$ 0.597 \pm 0.103 $ \\
SemCKD \citep{wang2022semckd} &	PANDA     &	$ 71.4 \pm 0.7 $ &	$ 0.644 \pm 0.010 $ &	$ 0.638 \pm 0.011 $ &	$ 79.0 \pm 0.9 $ &	$ 0.645 \pm 0.011 $ &	$ 0.700 \pm 0.013 $ \\
MoMA (Ours) &	ImageNet              &	$ 68.1 \pm 1.9 $ &	$ 0.611 \pm 0.036 $ &	$ 0.598 \pm 0.038 $ &	$ 68.1 \pm 1.2 $ &	$ 0.550 \pm 0.026 $ &	$ 0.559 \pm 0.069 $ \\
MoMA (Ours) &	PANDA                 &	$ \textbf{71.7} \pm \textbf{ 1.1} $ &	$ 0.663 \pm 0.011 $ &	$ 0.654 \pm 0.011 $ &	$ 68.8 \pm 4.6 $ &	$ 0.559 \pm 0.040 $ &	$ 0.590 \pm 0.068 $ \\
\midrule
Vanilla KD \citep{hinton2015distilling}  &	PANDA &	$ 71.6 \pm 1.2 $ &	$ 0.662 \pm 0.013 $ &	$ \textbf{0.656} \pm \textbf{0.011} $ &	$ 72.4 \pm 4.7 $ &	$ 0.590 \pm 0.044 $ &	$ 0.638 \pm 0.046 $ \\
SimKD \citep{chen2022knowledge} &	PANDA         &	$ 66.1 \pm 0.4 $ &	$ 0.411 \pm 0.007 $ &	$ 0.411 \pm 0.012 $ &	$ 78.5 \pm 0.7 $ &	$ 0.564 \pm 0.006 $ &	$ 0.651 \pm 0.009 $ \\
KL+FitNet \citep{DBLP:journals/corr/RomeroBKCGB14}  &	PANDA &	$ 64.9 \pm 6.5 $ &	$ 0.554 \pm 0.047 $ &	$ 0.556 \pm 0.038 $ &	$ 43.8 \pm 18.6 $ &	$ 0.331 \pm 0.129 $ &	$ 0.252 \pm 0.203 $ \\
KL+AT \citep{komodakis2017paying}  &	PANDA &	$ 68.1 \pm 0.9 $ &	$ 0.573 \pm 0.013 $ &	$ 0.582 \pm 0.009 $ &	$ 82.0 \pm 1.0 $ &	$ \textbf{0.648} \pm \textbf{0.013} $ &	$ 0.729 \pm 0.015 $ \\
KL+CC \citep{peng2019correlation} &	PANDA         &	$ 69.0 \pm 1.1 $ &	$ 0.584 \pm 0.014 $ &	$ 0.584 \pm 0.013 $ &	$ 82.0 \pm 1.6 $ &	$ 0.642 \pm 0.022 $ &	$ 0.722 \pm 0.023 $ \\
KL+CRD \citep{Tian2020Contrastive} &	PANDA         &	$ 69.3 \pm 1.1 $ &	$ 0.594 \pm 0.010 $ &	$ 0.596 \pm 0.011 $ &	$ 81.2 \pm 1.7 $ &	$ 0.625 \pm 0.031 $ &	$ 0.716 \pm 0.029 $ \\
KL+SemCKD \citep{wang2022semckd} &	PANDA     &	$ 68.5 \pm 0.4 $ &	$ 0.580 \pm 0.012 $ &	$ 0.581 \pm 0.016 $ &	$ 81.1 \pm 1.3 $ &	$ 0.637 \pm 0.040 $ &	$ 0.721 \pm 0.024 $ \\
KL+MoMA (Ours) &	PANDA                 &	$ 71.0 \pm 0.2 $ &	$ 0.606 \pm 0.010 $ &	$ 0.594 \pm 0.006 $ &	$ \textbf{83.4} \pm \textbf{0.1} $ &	$ 0.635 \pm 0.007 $ &	$ \textbf{0.760} \pm \textbf{0.001} $ \\

    \bottomrule
    \end{tabular}
    \end{adjustbox}
    \end{center}
\end{table*}

\begin{table*}[t!]
    \begin{center}
    \caption{Results of same task distillation on 25\% training data. KL denotes the use of KL divergence loss.}
    \label{table:prostate_tma_25}
    \setlength{\tabcolsep}{6pt} % Default value: 6pt
	\renewcommand{\arraystretch}{1} % Default value: 1
	\begin{adjustbox}{width=\textwidth}
	\begin{tabular}{c|c|ccc|ccc}
% \cline{3-8}
\cline{3-8}
 
\multicolumn{2}{c}{} & & Prostate USZ (Test I) & &	& Prostate UBC (Test II)	 &	 \\
\toprule
Method &	Pretrained &	ACC($\%$) &	F1 &	$\kappa_w$ &	ACC($\%$) &	F1 &	$\kappa_w$ \\
     \toprule
TC$_{PANDA}$ &	ImageNet &	$ 63.4 $ &	$ 0.526 $ &	$ 0.531  $ &	$ 78.2  $ &	$ 0.580 $ &	$ 0.680 $ \\
FT &	None       &	$ 63.8 \pm 2.3 $ &	$ 0.503 \pm 0.046 $ &	$ 0.478 \pm 0.058 $ &	$ 21.1 \pm 11.5 $ &	$ 0.137 \pm 0.069 $ &	$ 0.054 \pm 0.038 $ \\
FT &	ImageNet   &	$ 69.0 \pm 2.4 $ &	$ 0.612 \pm 0.033 $ &	$ 0.602 \pm 0.035 $ &	$ 70.4 \pm 2.2 $ &	$ 0.557 \pm 0.046 $ &	$ 0.541 \pm 0.079 $ \\
FT &	PANDA      &	$ 71.2 \pm 0.9 $ &	$ 0.641 \pm 0.008 $ &	$ 0.635 \pm 0.013 $ &	$ 71.3 \pm 4.1 $ &	$ 0.573 \pm 0.047 $ &	$ 0.588 \pm 0.075 $ \\
\midrule
FitNet \citep{DBLP:journals/corr/RomeroBKCGB14}  &	PANDA &	$ 65.2 \pm 5.7 $ &	$ 0.575 \pm 0.067 $ &	$ 0.564 \pm 0.075 $ &	$ 36.9 \pm 18.8 $ &	$ 0.273 \pm 0.138 $ &	$ 0.166 \pm 0.143 $ \\
AT \citep{komodakis2017paying}  &	PANDA &	$ 70.4 \pm 1.3 $ &	$ 0.636 \pm 0.020 $ &	$ 0.624 \pm 0.026 $ &	$ 76.8 \pm 1.7 $ &	$ 0.640 \pm 0.030 $ &	$ 0.667 \pm 0.023 $ \\
CC \citep{peng2019correlation}  &	PANDA         &	$ 72.4 \pm 0.8 $ &	$ 0.660 \pm 0.010 $ &	$ 0.653 \pm 0.008 $ &	$ 75.9 \pm 2.0 $ &	$ 0.620 \pm 0.025 $ &	$ 0.649 \pm 0.039 $ \\
CRD \citep{Tian2020Contrastive} &	PANDA     &	$ 70.2 \pm 1.3 $ &	$ 0.638 \pm 0.011 $ &	$ 0.628 \pm 0.010 $ &	$ 66.9 \pm 5.2 $ &	$ 0.546 \pm 0.039 $ &	$ 0.599 \pm 0.045 $ \\
SemCKD \citep{wang2022semckd} &	PANDA     &	$ 69.4 \pm 1.0 $ &	$ 0.622 \pm 0.014 $ &	$ 0.607 \pm 0.021 $ &	$ 73.9 \pm 2.5 $ &	$ 0.597 \pm 0.032 $ &	$ 0.608 \pm 0.059 $ \\
MoMA (Ours) &	ImageNet              &	$ 69.4 \pm 1.6 $ &	$ 0.630 \pm 0.025 $ &	$ 0.617 \pm 0.020 $ &	$ 68.8 \pm 2.8 $ &	$ 0.555 \pm 0.031 $ &	$ 0.579 \pm 0.037 $ \\
MoMA (Ours) &	PANDA                 &	$ \textbf{73.0} \pm \textbf{1.3} $ &	$ \textbf{0.660} \pm \textbf{0.013} $ &	$ \textbf{0.653} \pm \textbf{0.016} $ &	$ 73.7 \pm 3.6 $ &	$ 0.591 \pm 0.044 $ &	$ 0.582 \pm 0.067 $ \\
\midrule
Vanilla KD \citep{hinton2015distilling}  &	PANDA &	$ 72.0 \pm 0.3 $ &	$ 0.648 \pm 0.008 $ &	$ 0.638 \pm 0.008 $ &	$ 72.6 \pm 3.3 $ &	$ 0.589 \pm 0.035 $ &	$ 0.591 \pm 0.066 $ \\
SimKD \citep{chen2022knowledge} &	PANDA         &	$ 65.6 \pm 0.5 $ &	$ 0.402 \pm 0.008 $ &	$ 0.396 \pm 0.016 $ &	$ 78.2 \pm 0.7 $ &	$ 0.562 \pm 0.006 $ &	$ 0.644 \pm 0.013 $ \\
KL+FitNet \citep{DBLP:journals/corr/RomeroBKCGB14} &	PANDA &	$ 67.1 \pm 0.7 $ &	$ 0.571 \pm 0.016 $ &	$ 0.576 \pm 0.017 $ &	$ 61.3 \pm 12.2 $ &	$ 0.447 \pm 0.096 $ &	$ 0.396 \pm 0.177 $ \\
KL+AT \citep{komodakis2017paying}  &	PANDA &	$ 67.7 \pm 1.0 $ &	$ 0.574 \pm 0.015 $ &	$ 0.584 \pm 0.015 $ &	$ 80.9 \pm 1.0 $ &	$ 0.636 \pm 0.014 $ &	$ 0.709 \pm 0.023 $ \\
KL+CC \citep{peng2019correlation}  &	PANDA         &	$ 69.6 \pm 0.8 $ &	$ 0.606 \pm 0.007 $ &	$ 0.605 \pm 0.007 $ &	$ 81.6 \pm 0.8 $ &	$ 0.636 \pm 0.014 $ &	$ 0.722 \pm 0.010 $ \\
KL+CRD T\citep{Tian2020Contrastive} &	PANDA         &	$ 69.6 \pm 0.6 $ &	$ 0.599 \pm 0.009 $ &	$ 0.589 \pm 0.009 $ &	$ 81.2 \pm 1.3 $ &	$ 0.636 \pm 0.023 $ &	$ 0.706 \pm 0.031 $ \\
KL+SemCKD \citep{wang2022semckd} &	PANDA     &	$ 68.8 \pm 0.5 $ &	$ 0.588 \pm 0.005 $ &	$ 0.590 \pm 0.005 $ &	$ 82.0 \pm 0.4 $ &	$ \textbf{0.651} \pm \textbf{0.007} $ &	$ 0.734 \pm 0.007 $ \\
KL+MoMA (Ours) &	PANDA                 &	$ 68.8 \pm 1.4 $ &	$ 0.569 \pm 0.011 $ &	$ 0.569 \pm 0.01 $ &	$ \textbf{82.6} \pm \textbf{ 0.8} $ &	$ 0.607 \pm 0.009 $ &	$ \textbf{0.746} \pm \textbf{0.013} $ \\

    \bottomrule
    \end{tabular}
    \end{adjustbox}
    \end{center}
\end{table*}

\begin{table*}[!t]
    \begin{center}
    \caption{Results of irrelevant task distillation: prostate cancer classification.}
    \label{table:prostate_colonk19}
    \setlength{\tabcolsep}{6pt} % Default value: 6pt
	\renewcommand{\arraystretch}{1} % Default value: 1
	\begin{adjustbox}{width=\textwidth}
	\begin{tabular}{c|c|ccc|ccc}
% \cline{3-8}
\cline{3-8}
 
     % \toprule
\multicolumn{2}{c}{} & & Prostate USZ (Test I) & &	& Prostate UBC (Test II)	 &	 \\
\toprule
Method &	Pretrained &	ACC($\%$) &	F1 &	$\kappa_w$ &	ACC($\%$) &	F1 &	$\kappa_w$ \\
     \toprule

FT & Colon K19            & $ 65.8 \pm 3.6 $ &	$ 0.612 \pm 0.046 $ &	$ 0.593 \pm 0.036 $ &	$ 66.4 \pm 7.8 $ &	$ 0.549 \pm 0.071 $ &	$ 0.553 \pm 0.100 $ 	\\
MoMA (Ours) & 	Colon K19  & 	$ \textbf{68.4} \pm \textbf{2.2} $ &	$ \textbf{0.621} \pm \textbf{0.015} $ &	$ \textbf{0.599} \pm \textbf{ 0.008} $ &	$ \textbf{73.3} \pm\textbf{ 5.8} $ &	$ \textbf{0.624} \pm \textbf{0.063} $ &	$ \textbf{0.638} \pm \textbf{0.052} $ \\

    \bottomrule
    \end{tabular}
    \end{adjustbox}
    \end{center}
	\end{table*}

\section*{Appendix C. Confusion Matrices of Experimental Results.} 
\setcounter{figure}{0}
\renewcommand{\thefigure}{C\arabic{figure}}

For each of the test datasets, we provide the confusion matrices of all models. Each confusion matrix represents the average across 5 runs or 5-fold cross-validation experiments. Consequently, the total sample count for each class in the confusion matrix may differ slightly from the actual sample count in the dataset.

\begin{figure*}[b!]
	\centering
	\includegraphics[width=1.0\textwidth]{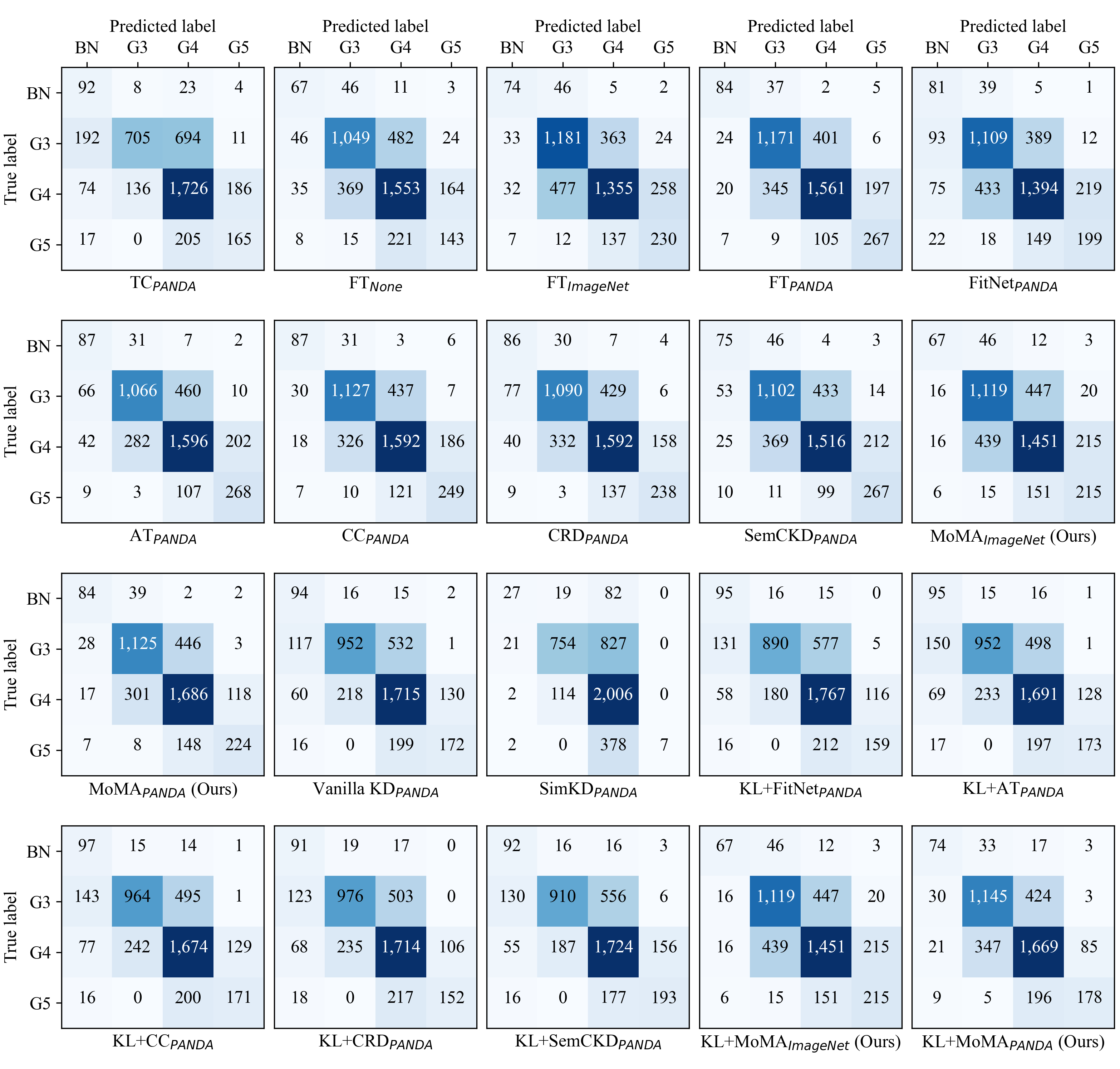}
	\caption{Confusion matrices on Prostate USZ (Test I). Each confusion matrix represents the average across 5 runs.}
	\label{fig:confusion_matric_number_prostate_1}
\end{figure*}

\begin{figure*}[b!]
	\centering
	\includegraphics[width=1.0\textwidth]{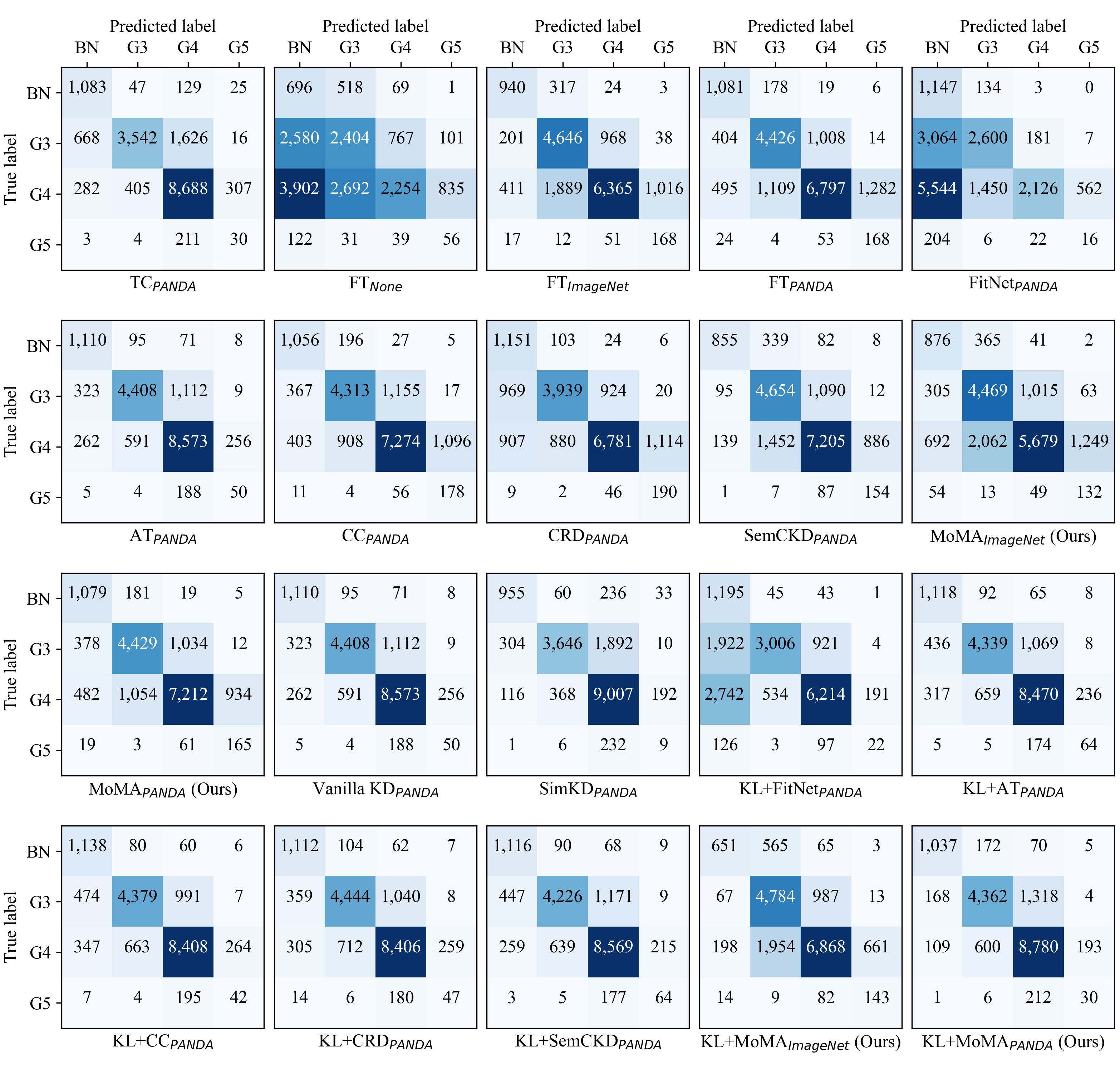}
	\caption{Confusion matrices on Prostate UBC (Test II). Each confusion matrix represents the average across 5 runs.}
	\label{fig:confusion_matric_number_prostate_2}
\end{figure*}

\begin{figure*}[b!]
	\centering
	\includegraphics[width=1.0\textwidth]{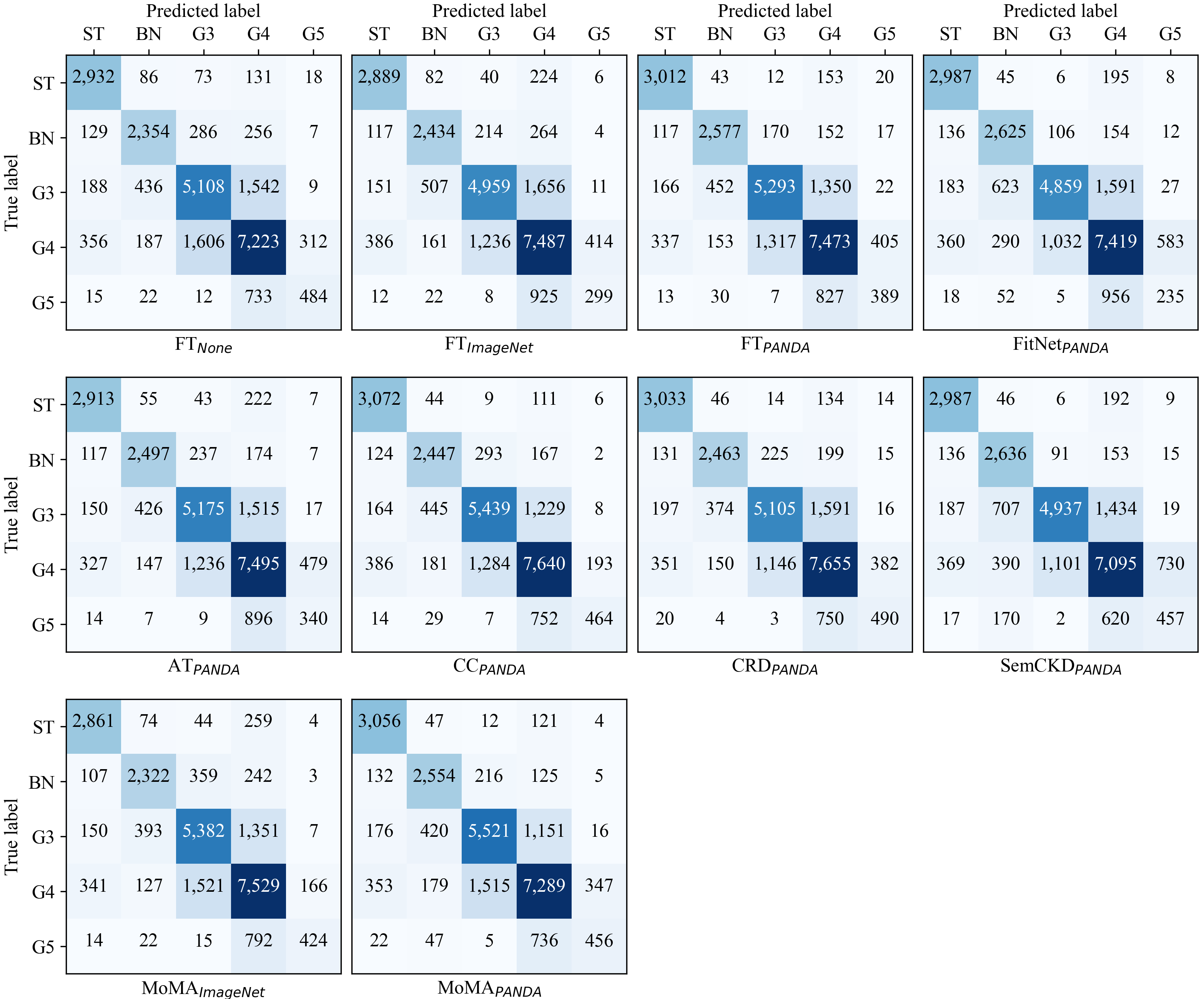}
	\caption{Confusion matrices on Prostate AGGC CV (Test I). Each confusion matrix represents the average across 5-fold cross-validation experiments.}
	\label{fig:confusion_matric_number_aggc_1}
\end{figure*}

\begin{figure*}[b!]
	\centering
	\includegraphics[width=1.0\textwidth]{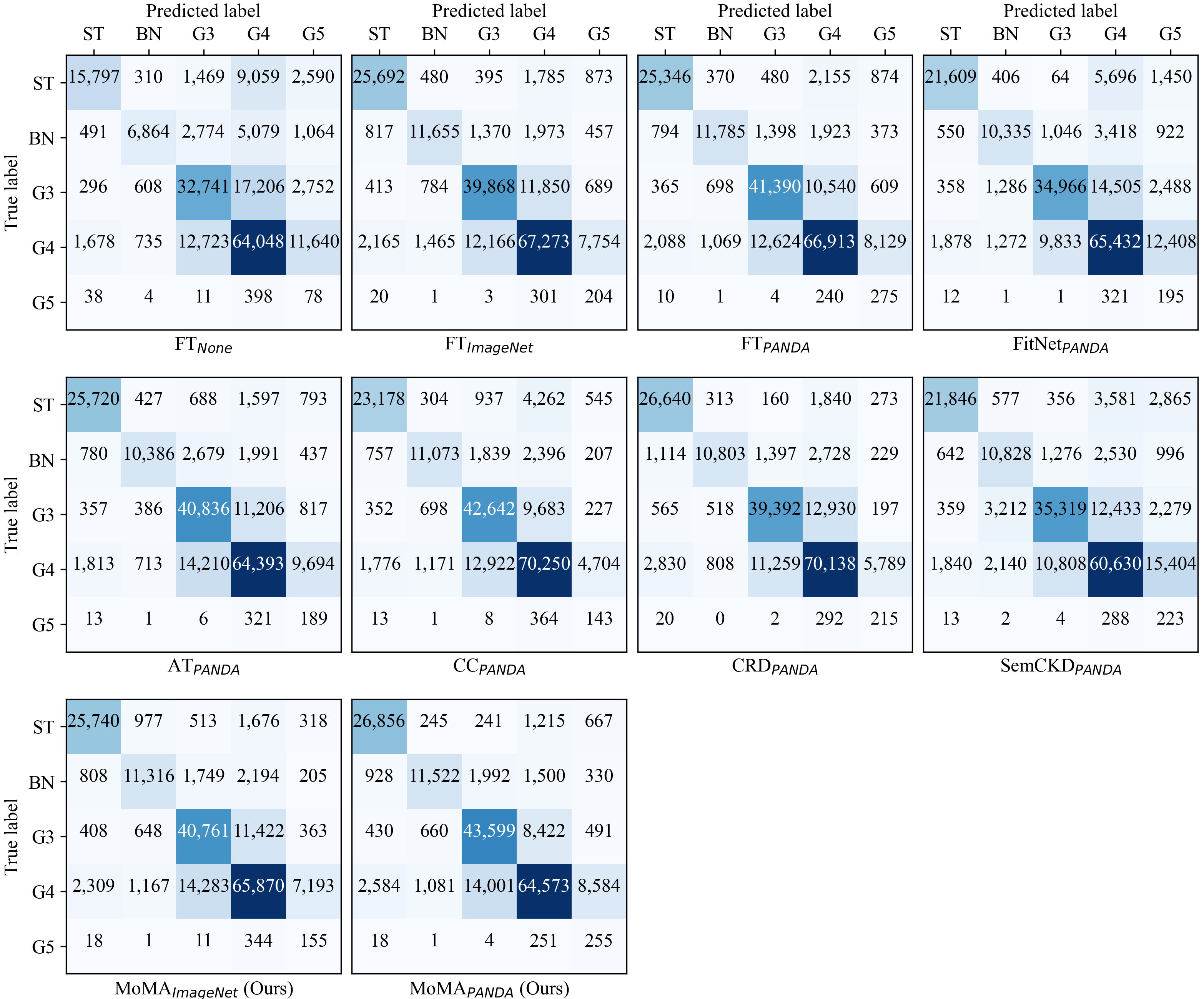}
	\caption{Confusion matrices on Prostate AGGC test (Test II). Each confusion matrix represents the average across 5-fold cross-validation experiments.}
	\label{fig:confusion_matric_number_aggc_2}
\end{figure*}
\begin{figure*}[b!]
	\centering
	\includegraphics[width=1.0\textwidth]{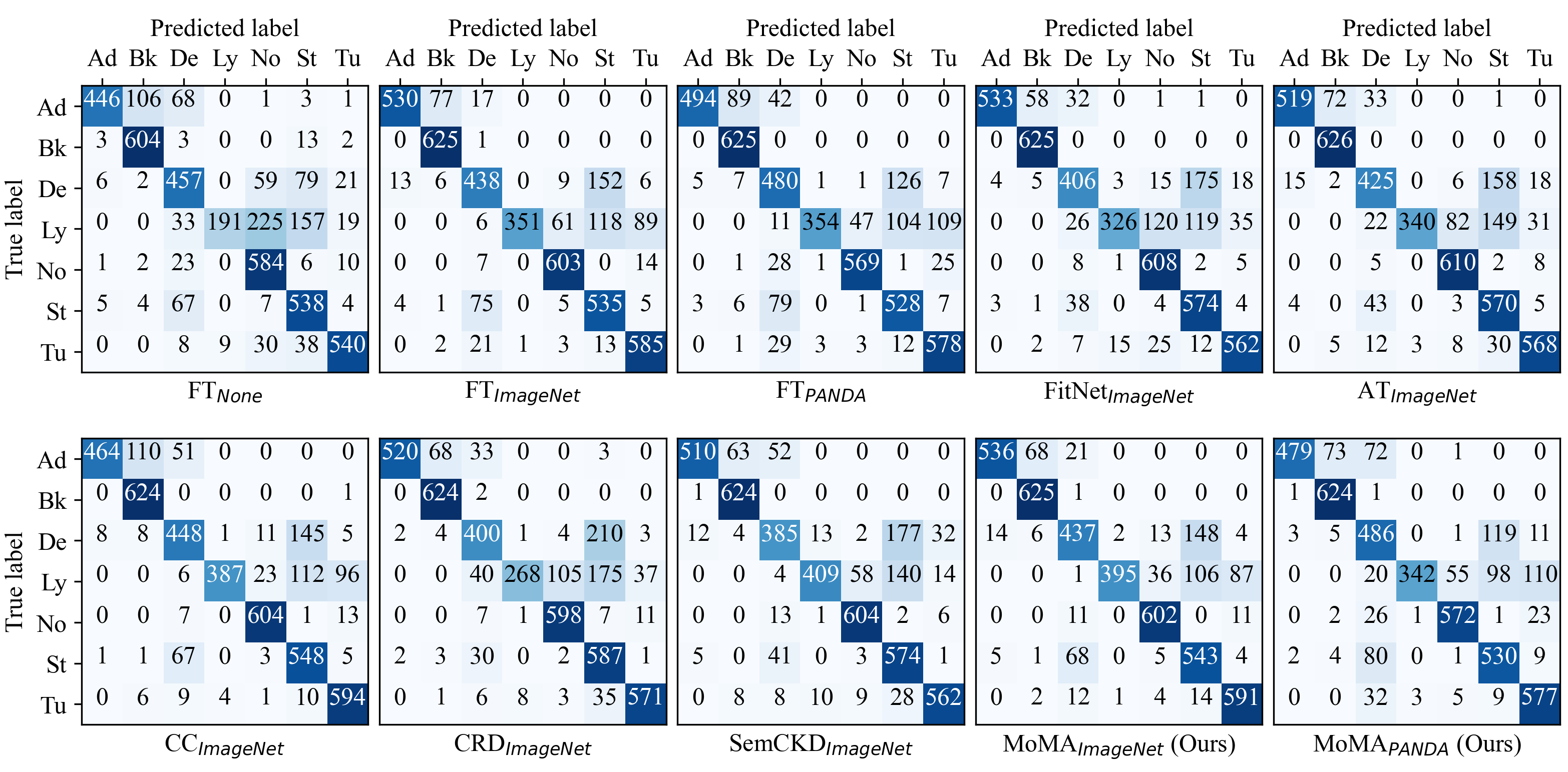}
	\caption{Confusion matrices on Colon K16 SN (Test I). Each confusion matrix represents the average across 5 runs.}
	\label{fig:confusion_matric_number_colon_1}
\end{figure*}
\begin{figure*}[b!]
	\centering
	\includegraphics[width=1.0\textwidth]{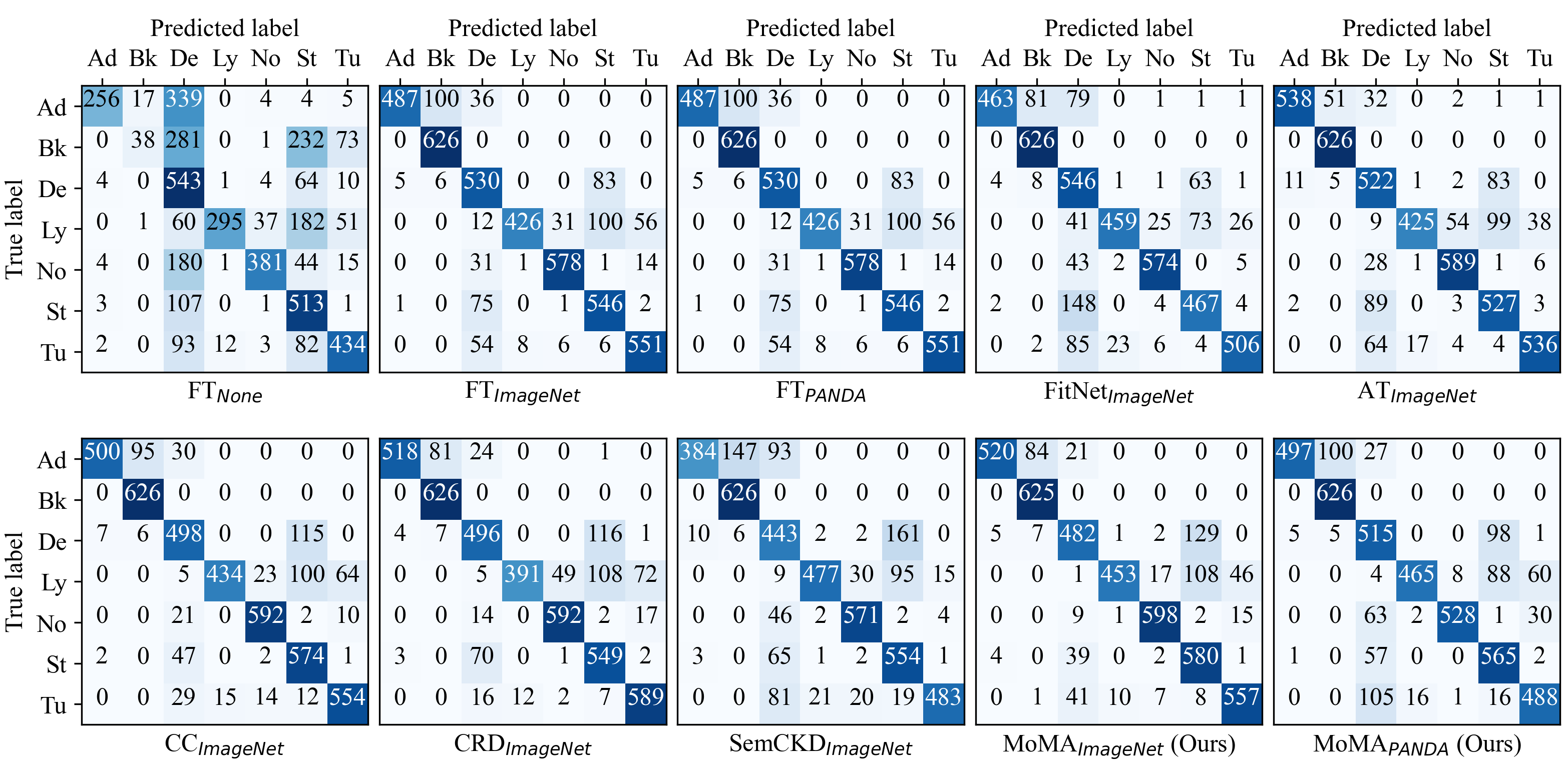}
	\caption{Confusion matrices on Colon K16 (Test II). Each confusion matrix represents the average across 5 runs.}
	\label{fig:confusion_matric_number_colon_2}
\end{figure*}

\begin{figure*}[b!]
	\centering
	\includegraphics[width=1.0\textwidth]{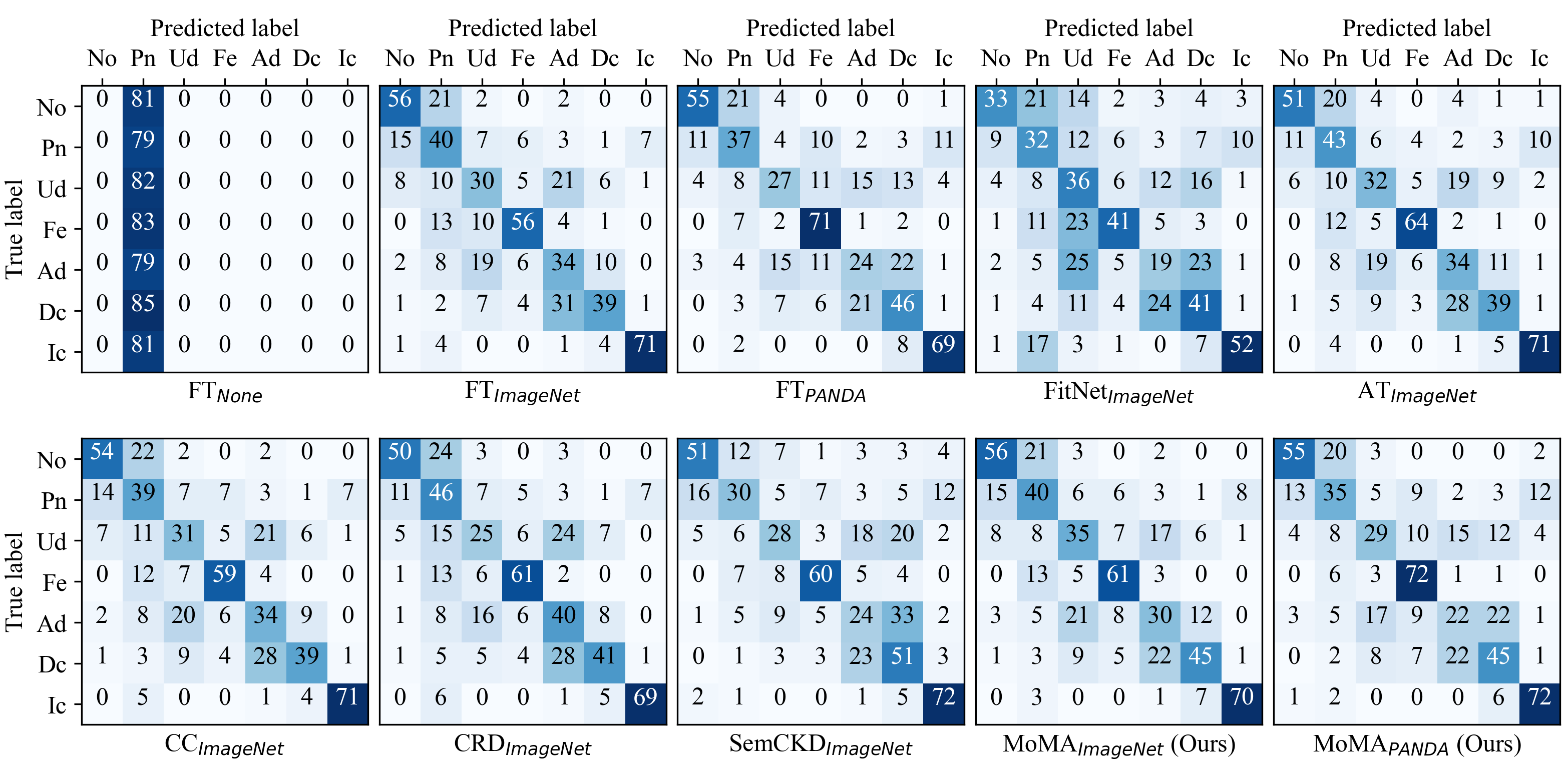}
	\caption{Confusion matrices on irrelevant task distillation: breast carcinoma sub-type classification. Each confusion matrix represents the average across 5 runs.}
	\label{fig:confusion_matric_number_breast}
\end{figure*}

\begin{figure*}[b!]
	\centering
	\includegraphics[width=1.0\textwidth]{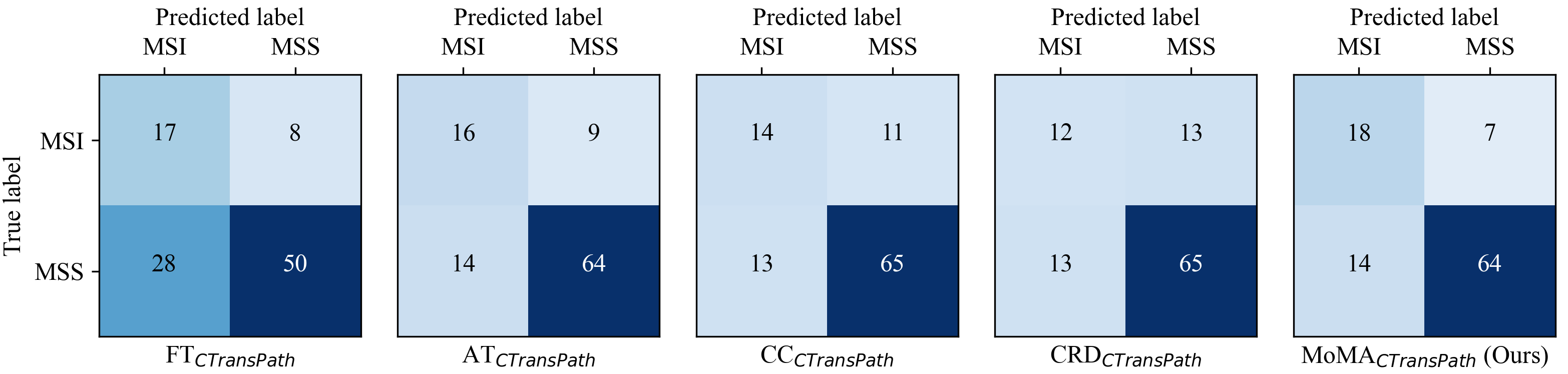}
	\caption{Confusion matrices on irrelevant task distillation: gastric microsatellite instability prediction. Each confusion matrix represents the average across 5-fold cross-validation experiments.}
	\label{fig:confusion_matric_number_gastric}
\end{figure*}

\end{document}